\documentstyle[12pt]{article}
\begin{document}
\begin{titlepage}
\begin{center}

\hfill YITP-SB-01/26  \\
\hfill {\tt hep-th/0106023}
\vskip 20mm

{\Huge Four dimensional ${\cal N}=1$ \\
\vskip 2mm
supersymmetrization of
${\cal R}^4$ in  superspace}

\vskip 10mm

Filipe Moura

\vskip 4mm

{\em C. N. Yang Institute for Theoretical Physics \\
State University of New York \\
Stony Brook, NY 11794-3840, U.S.A}

{\tt fmoura@insti.physics.sunysb.edu}

\vskip 6mm

\end{center}

\vskip .2in

\begin{center} {\bf Abstract } \end{center}
\begin{quotation}\noindent
We write an action, in four dimensional ${\cal N}=1$ curved superspace, 
which contains a pure ${\cal R}^4$ term with a coupling constant. Starting from
the off-shell solution of 
the Bianchi identities, we compute the on-shell torsions and curvatures with 
this term. We show that their complete solution includes, for some of them, an 
infinite series in the ${\cal R}^4$ coupling constant, which can only be 
computed iteratively. We explicitly compute the superspace torsions and 
curvatures up to second order in this coupling constant. Finally, we comment on
the lifting of this result to higher dimensions.
\end{quotation}

\vfill


\end{titlepage}

\eject


\section{Introduction}
\indent

The supersymmetrization of higher-derivative terms has been object of research
for a long time. Green and Schwarz proved that, in order to eliminate anomalies
from type I supergravity coupled to super Yang-Mills theories, one needs 
the Green-Schwarz mechanism \cite{gs84, gs85}, by which the supergravity two-form field strength is modified with the subtraction of the Lorentz Chern-Simons term. This mechanism violates local supersymmetry and originates an action with a ${\cal R}^2$ term.

In order to cancel the anomalies, it is also necessary to introduce bosonic 
higher-derivative counterterms (which include, for instance, ${\cal R}^4$ terms to cancel gravitational anomalies).
These counterterms have been shown to originate naturally in string theory
\cite{gs85}. They should then be part of the string theory low energy effective action, but they have to be introduced by hand on the supergravity action. Obviously they also break supersymmetry. 

Purely ``stringy'' ${\cal R}^4$ terms also show up in the low energy field 
theory effective action of both type II and heterotic string theories, as was 
shown
in \cite{gw86, gs87} by computing four graviton scattering amplitudes and in 
\cite{gz86} by calculating loop corrections to supersymmetric sigma models 
(the requirement that their $\beta$ function vanishes determines the equations 
of motion of bosonic background fields, from which one determines the effective
action).

The presence of those terms in M-theory has also been proven by
one- and two-loop superparticle scattering calculations. The complete ${\cal R}^4$ term arises from four-graviton scattering in eleven-dimensional supergravity. The quantum contributions to this process were analyzed at one-loop in eleven-dimensional supergravity compactified on $S^1$ \cite{rt97} and $T^2$
\cite{rt97, ggv97}. The latest results were extended to two loops in \cite{gkv99}. 

In the three cases we have described, we have ${\cal R}^2$ and ${\cal R}^4$ terms which are not supersymmetric. The requirement of having a supersymmetric anomaly-free effective action motivates the supersymmetrization of these terms.

The supersymmetrization of the Lorentz Chern-Simons terms has already been made \cite{bss87, bd89, bpt87, rrz89} (see also the review \cite{fp91} for a more complete list of references); the corresponding result for the ${\cal R}^4$ terms in string and M-theory effective actions is still under study, though some results exist both in ten \cite{bd89, dsw92} and eleven dimensions \cite{pvw00, cgnn00}. In reference \cite{dsw92} there is a systematic discussion of the different types of ${\cal R}^4$ superinvariants one may have.

All these results and claims should also be valid in four dimensions, when one dimensionally reduces from ten or eleven. It is useful to consider the same 
problems in four-dimensional ${\cal N}=1$ supergravity. What makes these problems easier, in principle, in this last case, is the 
knowledge of auxiliary fields, and the existence of a completely off-shell 
formulation of the theory in superspace. Also, the solution of these problems provides important information and consistency tests for the higher dimensional problems.

The coupling of the Lorentz Chern-Simons term to four-dimensional ${\cal N}=1$ 
supergravity consistent with supersymmetry and dimensional reduction from string theory has already been worked out \cite{cfgp85, cfv87, dfmp89}. In these references, a linear multiplet is coupled to supergravity. The scalar belonging to that multiplet plays the role of the dilaton in the dimensionally reduced theory; the two-index gauge field in the same multiplet is also analogous to the ten-dimensional two-form whose field strength is corrected the Lorentz Chern-Simons term.

The problem of supersymmetrizing ${\cal R}^4$ in ${\cal N}=1$ four-dimensional 
supergravity had never 
been worked out.  From what has been said, this problem reduces then to simply 
writing the appropriate action in superspace and deriving the respective 
torsions and curvatures. That is what is done in this paper. In a future paper,
we will present the complete action in $x$ space. 

In any case, already from the results of this paper we can conclude that
it requires an infinite number of terms to achieve the complete 
supersymmetrization (after the elimination of auxiliary fields), i.e. to go
``on-shell''. The same is valid for writing the superspace torsions and 
curvatures. Those terms are part of an infinite series in the  ${\cal R}^4$ coupling constant, which cannot be written explicitly. In this paper, we compute the superspace torsions and 
curvatures up to second order in this coupling constant.

In section \ref{2} we review briefly how to derive field equations in 
superspace for pure ${\cal N}=1, d=4$ supergravity.

In section \ref{3} we motivate and write the superspace action containing the
${\cal R}^4$ term.

In section \ref{4} we determine the field equations from this action. We show 
that we have two field equations, the solutions of which being respectively a 
polynomial and an infinite series.

Finally, in section \ref{5} we compare this case with other 
higher-derivative superinvariants known in the literature, and we comment on 
the implications of our result for the determination of effective actions for
superstring/M theory.

In appendix \ref{appendix1} we present our choice of conventions. 


\section{A review of pure ${\cal N}=1, d=4$ supergravity in superspace}
\label{2}
\indent

In this section, we make a brief review of the superspace formulation of pure 
${\cal N}=1,$ $d=4$ supergravity, with emphasis on the process of deriving the 
field equations. None of the results mentioned in this section is original;
we just include them here for the reader to be acquainted with them
(written in our conventions), and because they are essencial for the rest of 
the paper, which is a generalization of these results to a superspace action
including the ${\cal R}^4$ terms. 

What is special about ${\cal N}=1, d=4$ supergravity is the existence of a 
completely off-shell formalism. This means that a complete set of auxiliary 
fields is known (actually, there exist three known choices). In superspace 
this means that, after imposing constraints on the torsions, we can 
completely solve the Bianchi identities without using the field equations
\cite{gwz79}. Our choice of constraints on the torsions and
the solutions to the Bianchi identities are listed, in our conventions, in 
appendix \ref{appendix1.2}. The main result is that we can express all the 
unconstrained torsions and curvatures as functions of three superfields $R$, 
$G_{A \dot A}$ and $W_{ABC}$ (and their complex conjugates). These superfields
have some properties and satisfy some differential constraints also listed in 
appendix \ref{appendix1.2}.

Another special feature of pure ${\cal N}=1$ four-dimensional supergravity is 
that its action in superspace is known. It is written as the integral, over 
the whole superspace, of the superdeterminant of the supervielbein 
\cite{wz781}:

\begin{equation}
I=\frac{1}{2 \kappa^2} \int \int E d^4\theta d^4x , 
E=\mbox{sdet}E_\Lambda^{\ \ M} \label{pure}
\end{equation}

The following well known result is very useful when computing field equations 
in superspace from a given action:

\begin{equation}
\int E\nabla_M v^M\left( -\right)^M d^4x d^4\theta =0 \label{div}
\end{equation}
It allows us to integrate by parts and discard the terms in the action which 
are full divergences. This result follows uniquely from the torsion 
constraints and the Bianchi identities. 

Some of the constraints on the torsions allow us to express all spin 
connections in terms of supervielbeins. The other torsion constraints 
represent constraints on the supervielbein. One can express 
$E_m^{\ \ \Lambda }$ in terms of $E_A^{\ \ \Lambda}$ and 
$E_{\dot A}^{\ \ \Lambda}$, which can be expressed in terms of prepotentials 
\cite{gs80}. But these constraints do not depend on any action, and should be
preserved when we vary the supervielbeins and superconnections, in order to
derive the field equations for any action we take. Therefore we define, 
according to the original work \cite{wz781} and the review \cite{z78},
\begin{equation}
H_M^{\ \ N}=E_M^{\ \ \Lambda } \delta E_\Lambda^{\ \ N} \Rightarrow \delta
E_\Lambda^{\ \ N}=E_\Lambda^{\ \ M} H_M^{\ \ N} \label{hmn}
\end{equation}
\begin{equation}
\Phi_{MN}^{\ \ \ \ P} =E_M^{\ \ \Lambda } \delta \Omega_{\Lambda N}^{\
\ \ \ P} \label{phimn}
\end{equation}
as arbitrary variations of the supervielbein and the superconnection, 
respectively, subject to the condition that the torsion constraints should 
remain satisfied. $\Phi_{MN}^{\ \ \ \ P}$ is Lorentz-valued, like 
$\Omega_{M N}^{\ \ \ \ P}$:
$\Phi_{MB}^{\ \ \ \ C}=-\frac{1}{4} \Phi_{Mmn} \left( \sigma^{mn}
\right)_B^{\ \ C} = \frac{1}{4} \Phi_{M B \dot B}^{\ \ \ \ \ \ C \dot B}.$

From the constrained variations (\ref{hmn}), (\ref{phimn}) and the definition 
(\ref{tor}) of torsion, we derive the constrained variation
\begin{eqnarray} 
\delta T_{MN}^{\ \ \ \ R} &=& -H_M^{\ \ S} T_{SN}^{\ \ \ R}
+\left(-\right)^{MN} H_N^{\ \ S} T_{SM}^{\ \ \ R} +T_{MN}^{\ \
\ \ S} H_S^{\ \ R} \nonumber \\
&-&\nabla_M H_N^{\ \ R} +\left(-\right)^{MN} \nabla_N
H_M^{\ \ R} +\Phi_{MN}^{\ \ \ \ R}-\left(-\right)^{MN} \Phi_{NM}^{\ \
\ \ R} \label{deltatmnr}
\end{eqnarray}

The equations for $\delta T_{MN}^{\ \ \ R}$ are invariant under the
following two gauge transformations:

\begin{eqnarray} 
\delta H_M^{\ \ \ N} &=& \nabla_M \xi^N- \xi^P T_{PM}^{\ \ \ \ N}
 \nonumber \\
\delta \Phi_{MN}^{\ \ \ \ P} &=& \xi^Q R_{QMN}^{\ \ \ \ \ \ P} \label{gauge1}
\end{eqnarray}
and
\begin{eqnarray}
\delta H_M^{\ \ N} &=& X_M^{\ \ N}
\nonumber \\
\delta \Phi_{MN}^{\ \ \ \ P} &=& \nabla _M X_N^{\ \ P} \label{gauge2}
\end{eqnarray}

The method for deriving, from a given action, its field equations in 
superspace has been exposed in \cite{wz781, z78, west, bgg01}. For 
completeness, we summarize it here.
The main idea is to determine the whole set of $H_M^{\ \ N}$ and 
$\Phi_{MN}^{\ \ \ \ P}$. This can be achieved by applying (\ref{deltatmnr}) to 
the constrained torsions. From what we mentioned before, their variation should
be zero, which implies a set of algebraic equations from which $H_M^{\ \ N}$ 
and $\Phi_{MN}^{\ \ \ \ P}$ can be determined in terms of the unconstrained 
supertorsions and supercurvatures (which means in terms of $R$, $G_{A \dot A}$ 
and $W_{ABC}$ and their complex conjugates), and some extra arbitrary 
superfields. Since these extra superfields are arbitrary, $\delta I$ can only 
vanish if their coefficients vanish. The vanishing of these coefficients is 
equivalent to the superspace field equations.

We will briefly review the pure supergravity case, before going to the
${\cal R}^4$ action. Details and derivations can be found in 
\cite{wz781, z78, west, bgg01}. The variation of the action (\ref{pure}) is 
simply given by
\begin{equation} 
2 \kappa^2
\delta I=\int E \left( \frac{1}{2} H_{A \dot A}^{\ \ \ A \dot A} -H_A^{\ \
A}-H_{\dot A}^{\ \ \dot A} \right) d^4x d^4\theta \label{deltapure}
\end{equation}

From $\delta T_{AB}^{\ \ \ \ \dot C}=0,\, \delta T_{AB}^{\ \ \ \
m}=0,$ and by fixing the gauge invariance (\ref{gauge1}), it has been shown
that we may write

\begin{eqnarray} 
H_{\dot B k}&=&-\nabla _{\dot B} \chi _k, \, \, 
H_{Bk}=\nabla _B \chi _k \label{hbk}
\\
H_{\dot B A}&=&\frac{i}{24} \overline{R} \chi_{A\dot B},  \, \, 
H_{A\dot B}=-\frac{i}{24} R \chi_{A \dot B} \label{hadb}
\end{eqnarray}
$\chi _k$ is an arbitrary, {\it pure imaginary} superfield.

It has also been shown that we have the 
following parameterization \footnote{Equations (\ref{hbk}), (\ref{hadb}), 
(\ref{hkkhaa}) and (\ref{hkkhdada}) are derived in \cite{west, bgg01} and were told to us by P. van Nieuwenhuizen.}:

\begin{eqnarray} 
H_k^{\ k} - H_A^{\ A} +\nabla^k \chi_k -2i \chi^k G_k &=& \left(
\overline{\nabla}^2 +\frac{1}{3} \overline{R} \right) \overline{U}
\label{hkkhaa} \\
H_k^{\ k} - H_{\dot A}^{\ \dot A} -\nabla^k \chi_k -2i \chi^k G_k
&=& \left(\nabla^2 +\frac{1}{3} R \right) U \label{hkkhdada}
\end{eqnarray}
$\left(\nabla^2 +\frac{1}{3} R \right)$ is, in our notations, the chiral 
projector for a scalar superfield. $U$ is an arbitrary superfield, and 
$\overline{U}$ its complex conjugate.

The final result for (\ref{deltapure}) is \cite{z78} \footnote{We will present 
our derivation of (\ref{deltapure3}) in section \ref{4}, after we have
computed the remaining components of $H_M^{\ \ N}$.}
\begin{eqnarray}
H_k^{\ k} - H_A^{\ A} - H_{\dot A}^{\ \dot A} 
&=& \frac{i}{12}\left( \nabla^{\dot
A}\nabla^A-\nabla^A\nabla^{\dot A}\right) \chi_{A\dot
A}+\frac{2}{3}i\chi^{A\dot A}G_{A\dot A} \nonumber \\
&+& \frac{1}{3}\left( \overline{\nabla}^2 
+\frac{1}{3}\overline{R}\right) \overline{U}
+\frac{1}{3}\left(\nabla^2+\frac{1}{3} R\right) U \label{deltapure3}
\end{eqnarray}
Replacing in (\ref{deltapure}), and recalling (\ref{div}),
\begin{equation}
2 \kappa^2
\delta I=\int E \left( \frac{2}{3} i\chi^{A \dot A} G_{A \dot A}
+\frac{1}{9} \overline{R} \overline{U} + \frac{1}{9} R U \right) d^4x
d^4\theta =0 \label{deltapure2}
\end{equation}

Since $U, \chi^{A \dot A}$ are completely arbitrary, there's no other 
possibility than to have 

\begin{eqnarray}
G_{A \dot A}&=&0 \label{g=0}
\\
R&=&0 \label{r=0}
\end{eqnarray}
These are the field equations for pure ${\cal N}=1$ four-dimensional 
supergravity. 


\section{The supersymmetric ${\cal R}^4$ action}
\label{3}
\setcounter{equation}{0}
\indent

To the action (\ref{pure}), we are adding (supersymmetric) ${\cal R}^4$
correction terms. Because of the field equations (\ref{g=0}), (\ref{r=0})
it does not make sense to add terms to (\ref{pure}) which are proportional
to $R$ or $G_{A \dot A}$ (the field equations would only get perturbative corrections which would
not affect the unperturbed solutions (\ref{g=0}), (\ref{r=0})). Therefore, the
terms we are looking for must contain $W_{ABC}$ and $W_{\dot A \dot B \dot C}$.
Indeed, it is well known \cite{ggrs} that the 
component expansion of $W_{ABC}$ includes the term \footnote{The underlined 
indices are symmetrized with weigth one; the vertical bar means the $\theta=0$
component.}
\begin{equation}
\left. \nabla_{\underline{A}} W_{\underline{BCD}} \right| = \frac{1}{8} 
{\cal R}_{\mu \nu \rho \sigma} \sigma^{\mu \nu}_{\underline{AB}} 
\sigma^{\rho \sigma}_{\underline{CD}} + \cdots 
= -\frac{1}{2}{\cal W}_{ABCD} + \cdots
\label{weyl}
\end{equation}
${\cal W}_{ABCD}$ is the Weyl tensor in spinor notation, a Lorentz-irreducible 
component of the Riemannn tensor \cite[chapters 4 and 13]{wald}.

One can easily (using the solutions to the Bianchi identities and 
(\ref{diffg}), (\ref{diffw})) derive the following formula:

\begin{eqnarray}
\nabla^2 W^2 &=& -2 \left(\nabla^A W^{BCD} \right) \nabla_{\underline{A}}
W_{\underline{BCD}} +12 \left(\nabla^m G^n \right) \nabla_m G_n \nonumber \\
&-& 12 \left(\nabla^m G^n \right) \nabla_n G_m -12i \varepsilon^{mnrs} 
\left(\nabla_m G_n \right) \nabla_r G_s  \nonumber \\
&-&\frac{5}{3} R W^2 - 20 W^{ABC} G_A^{\ \ \dot A} \nabla_B G_{C \dot A}
+8i W^{ABC} \nabla_A^{\ \ \dot A} \nabla_B G_{C \dot A} 
\label{d2w2}
\end{eqnarray}
We see that the $\theta^2$ component of $W^2$ has a ${\cal W}^2$ term. This 
way, a supersymmetric action which includes ${\cal R}^4$ (actually, 
${\cal W}^4$) should have (because of hermiticity) a $W^2 \overline{W}^2$ term. 
The action we are then considering \footnote{In a future publication \cite{m01}
we will discuss other possibilities for supersymmetric ${\cal R}^4$ actions.} 
is written, in superspace, in the following way:

\begin{equation}
I=\frac{1}{2 \kappa^2} \int \int E\left( 1+\alpha W^2\overline{W}^2\right) d^4\theta d^4x 
\label{action}
\end{equation}
$\alpha$ is a coupling constant, a perturbative parameter of mass dimension -6.
From now on, in each formula we write, the limit $\alpha=0$ is the limit of 
pure supergravity.

It is important to know the exact expansion in components of (\ref{action}). 
We leave the details for another publication
\cite{m01}. Anyway, it is useful to mention here some aspects. It is well known
that, in the old minimal formulation of supergravity \cite{sw78, fvn78}, there
are no fermionic auxiliary fields. The auxiliary fields are then a scalar $M$,
a pseudoscalar $N$ and a vector $A_m$, which are zero on-shell (in pure 
supergravity). The following identification is valid 
\cite{wz782}:
\begin{eqnarray}
\left. \overline{R}\right| &=& 4 \left( M+iN \right)
\\
\left. G_{A\dot A}\right| &=&\frac{1}{3} A_{A\dot A} \label{g=a}
\end{eqnarray}
It agrees with the field equations (\ref{g=0}), (\ref{r=0}) of pure 
supergravity. But if we consider our higher-derivative action (\ref{d2w2}), we
see that the $\theta^2$ component of $W^2$ has derivatives of $G_{A\dot A}$. 
If we expand then (\ref{action}) in components, we will also
get a higher-derivative action in the auxiliary field $A_m$, which means a 
complicated nonalgebraic field equation for it. 

It is worth mentioning that the piece of (\ref{action}) proportional to $\alpha$, 
i.e. the ${\cal R}^4$ correction to pure supergravity, is also the subleading 
correction to the Weyl supergravity action in a recently proposed 
Born-Infeld-Weyl supergravity action \cite{gk01}. 


\section{The superspace field equations, torsions and curvatures}
\label{4}
\setcounter{equation}{0}
\indent

Now we are ready to follow the same procedure as in sec. \ref{2}, but this time
with respect to the action (\ref{action}). We will need the whole set of 
components of $H_M^{\ \ N}$ and all but one of $\Phi_{MN}^{\ \ \ \ P}$. We 
start by this computation.

\subsection{The full computation of $H_M^{\ \ N}$ and $\Phi_{MN}^{\ \ \ \ P}$}
\indent

From the variation of the constrained torsions, we get

\begin{eqnarray} 
\delta T_{AB}^{\ \ \ C}=0 &\Rightarrow& -\frac{1}{2}
H_{\underline{A}}^{\ D \dot D} T_{D \dot D \ \underline{B}}^{\
\ \ \ \ \ C} -\nabla_{\underline{A}} H_{\underline{B}}^{\ \ C}
+\Phi_{\underline{A}\underline{B}}^{\ \ \ C}=0 
\\
\delta T_{A\dot B}^{\ \ \ \dot C}=0 &\Rightarrow&  
-\frac{1}{2} H_A^{\ D\dot D} T_{D\dot D \dot B}^{\ \ \ \ \ \dot C} 
-\frac{1}{2} H_{\dot B}^{\ D\dot D} T_{D \dot D A}^{\ \ \ \ \ \dot C}
+T_{A\dot B}^{\ \ \ n} H_n^{\ \dot C} 
\nonumber \\ 
&\left. \right.& -  \nabla _A H_{\dot B}^{\
\dot C} -\nabla_{\dot B} H_A^{\ \ \dot C} +\Phi_{A\dot B}^{\ \ \ \dot C}=0
\\ 
\delta T_{A \dot B}^{\ \ \ m}=0 &\Rightarrow& -H_A^{\ C} T_{C\dot
B}^{\ \ \ m} - H_{\dot B}^{\ \dot C} T_{\dot C A}^{\ \ m}
+T_{A \dot B}^{\ \ n} H_n^{\ m} 
\nonumber \\ 
&\left. \right.&  -  \nabla_A H_{\dot B}^{\ m} -\nabla_{\dot B} H_A^{\ m}=0
\\
\delta T_{Am}^{\ \ \ n}=0 &\Rightarrow& H_m^{\ \ \dot B} T_{\dot
BA}^{\ \ \ n} + T_{Am}^{\ \ \ B} H_B^{\ n} +T_{Am}^{\ \
\ \dot B} H_{\dot B}^{\ n} -\nabla _A H_m^{\ \ n} +\nabla_m H_A^{\ n}  
\nonumber \\
&\left. \right.& + \Phi_{Am}^{\ \ n}=0 
\\
\delta T_{mn}^{\ \ \ p}=0 &\Rightarrow& T_{mn}^{\ \ \ A} H_A^{\
p}+ T_{mn}^{\ \ \ \dot A} H_{\dot A}^{\ p} -\nabla_m H_n^{\ \ p}
+\nabla_n H_m^{\ \ p} 
 \nonumber \\ 
&\left. \right.& + \Phi_{mn}^{\ \ \ p} -\Phi_{nm}^{\ \ \ p}=0
\end{eqnarray}

From $\delta T_{Am}^{\ \ \ \ n}=0$, we get

\begin{eqnarray}
\Phi_{AB \dot B}^{\ \ \ \ \ \ C \dot C}&=&-H_{B \dot B}^{\ \ \ \dot
A} T_{\dot A A}^{\ \ \ C \dot C}- T_{AB\dot B}^{\ \ \ \ \ D}
H_D^{\ C\dot C}- T_{AB \dot B}^{\ \ \ \ \dot D} H_{\dot D}^{\ \ C
\dot C} \nonumber \\
&+& \nabla_A H_{B\dot B}^{\ \ \ C \dot C}- \nabla_{B \dot B}
H_A^{\ C \dot C} \label{Phiabdb}
\end{eqnarray}

From $\delta T_{mn}^{\ \ \ \ p}=0$ we get an algebraic equation for
$\Phi_{mn}^{\ \ \ \ p}$. This specific computation is not necessary for the 
variation of our action (see sec. \ref{42}); once we determine 
$H_m^{\ \ p}$, as we will, and knowing $H_A^{\ \ p}$, this computation becomes
straightforward.

From $\delta T_{A \dot B}^{\ \ \ \dot C}=0$ we get 

\begin{eqnarray} 
\Phi_{A \dot B}^{\ \ \ \ \dot C}
&=&\frac{1}{4} \Phi_{AB \dot B}^{\ \ \ \ \ \ B \dot C}  
=\frac{1}{2} H_A^{\ \ D \dot D} T_{D \dot D \dot B}^{\ \
\ \ \ \dot C} 
+\frac{1}{2} H_{\dot B}^{\ \ D \dot D} T_{D\dot D A}^{\ \ \ \ \ \dot C}
\nonumber \\
&+& 2i H_{A \dot B}^{\ \ \ \dot C}
+\nabla_A H_{\dot B}^{\ \ \dot C} 
+ \nabla_{\dot B} H_A^{\ \ \dot C}
\end{eqnarray}

From $\delta T_{A \dot B}^{\ \ \ \ m}=0$

\begin{equation} 
2iH_A^{\ \ C} \sigma_{C \dot B}^{\ \ \ m} +2i H_{\dot B}^{\ \ \dot C}
\sigma_{A \dot C}^{\ \ \ m} -2i H_{A \dot B}^{\ \ \ m} -\nabla_A
H_{\dot B}^{\ m} -\nabla_{\dot B} H_A^{\ m}=0
\end{equation} 
from which we get the equation

\begin{equation} 
H_{A \dot A}^{\ \ \ B \dot B}= 2\left( H_A^{\ \ B} \varepsilon_{\dot A}^{\
\ \dot B} +H_{\dot A}^{\ \ \dot B} \varepsilon_A^{\ \ B} \right) +\frac{i}{2}
\left( \nabla_A H_{\dot A}^{\ \ \ B \dot B} +\nabla_{\dot A} H_A^{\ \ \ B
\dot B} \right) \label{hadabdb}
\end{equation}

We can use the gauge invariance (\ref{gauge2}) to gauge away the symmetric 
part of $H_{AB}$:
\begin{eqnarray} 
H_A^{\ \ B} &=& \frac{1}{2} H_C^{\ \ C} \delta_A^{\ \ B}
\nonumber \\ 
H_{\dot A}^{\ \ \dot B} &=& \frac{1}{2} H_{\dot C}^{\ \ \dot C} \delta_{\dot
A}^{\ \ \dot B} \label{hab}
\end{eqnarray}
Replacing (\ref{hbk}) and (\ref{hab}) into (\ref{hadabdb}), we get

\begin{equation} 
H_{A\dot A}^{\ \ \ B\dot B} =\left( H_C^{\ \ C} +H_{\dot C}^{\ \ \dot
C}\right) \delta_A^{\ \ B} \delta_{\dot A}^{\ \ \dot B} +\frac{i}{2} \left(
\nabla_A \nabla_{\dot A} -\nabla_{\dot A} \nabla_A\right) \chi^{B \dot B}
\label{hadabdb2}
\end{equation}
Tracing the last equation for $H_{A \dot A}^{\ \ \ B \dot B}$ we obtain

\begin{equation} 
H_k^{\ k}= 2 \left( H_A^{\ \ A} +H_{\dot A}^{\ \ \dot A} \right)
-\frac{i}{4} \left( \nabla_A \nabla_{\dot A} -\nabla_{\dot A} \nabla_A
\right) \chi^{A \dot A}
\end{equation}
Combining this last equation with the previous equations (\ref{hkkhaa}) and
(\ref{hkkhdada}), we can solve for 

\begin{eqnarray} 
H_k^{\ k}&=&
\frac{8}{3}i \chi^k G_k
+ \frac{2}{3} \left( \overline{\nabla}^2
+\frac{1}{3} \overline{R} \right) \overline{U}
+ \frac{2}{3} \left( \nabla^2 +\frac{1}{3} R \right) U \nonumber \\
&+& \frac{i}{12} \left( \nabla_A \nabla_{\dot A}
-\nabla_{\dot A} \nabla_A \right) \chi^{A \dot A}
\end{eqnarray}

\begin{eqnarray} 
H_A^{\ \ A} &=&
\frac{2}{3}i \chi^k G_k 
-\frac{1}{3} \left( \overline{\nabla}^2 +\frac{1}{3} \overline{R} \right) 
\overline{U}
+ \frac{2}{3} \left( \nabla^2 +\frac{1}{3} R \right) U \nonumber \\
&+& \frac{i}{12} \left( \nabla_A \nabla_{\dot A} -\nabla_{\dot A} \nabla_A
\right) \chi^{A \dot A}
+\nabla^k \chi _k
\end{eqnarray}

\begin{eqnarray} 
H_{\dot A}^{\ \ \dot A}
&=&\frac{2}{3}i \chi^k G_k
+ \frac{2}{3} \left(\overline{\nabla}^2 + \frac{1}{3} \overline{R} \right) 
\overline{U}
-\frac{1}{3} \left( \nabla^2+ \frac{1}{3} R \right) U \nonumber \\
&+&\frac{i}{12} \left( \nabla_A \nabla_{\dot A} - \nabla_{\dot A} \nabla_A
\right) \chi^{A \dot A} - \nabla^k \chi_k
\end{eqnarray}
From these three expressions, (\ref{deltapure3}) follows. Also, from 
(\ref{hadabdb2}), it follows that
\begin{eqnarray} 
H_{A \dot A}^{\ \ \ B \dot B} &=& 
\left[ \frac{1}{3} \left(\nabla^2 
+\frac{1}{3} R \right) U
+\frac{1}{3} \left( \overline{\nabla}^2 
+\frac{1}{3} \overline{R} \right) \overline{U} 
+\frac{2}{3} i \chi^{C \dot C} G_{C \dot C} \right. \nonumber \\
&+& \frac{i}{6} \left( \nabla_C \nabla_{\dot C} -\nabla_{\dot C}
\nabla_C \right) \chi^{C \dot C} \left.  \right]
\varepsilon_A^{\ \ B} \varepsilon_{\dot A}^{\ \ \dot B} \nonumber \\
&+& \frac{i}{2} \left( \nabla_{\dot A} \nabla_A -\nabla_A
\nabla_{\dot A} \right) \chi^{B \dot B}
\end{eqnarray}
Replacing these expressions in (\ref{Phiabdb}) - actually its complex 
conjugate-, we can now solve for $H_{A \dot A}^{\ \ \ B}$:

\begin{eqnarray} 
H_{A\dot A}^{\ \ \ B} &=& 
\frac{1}{6} \varepsilon_A^{\ \ B} \nabla_{\dot A} \left( \chi^{D \dot D} G_{D 
\dot D} \right) 
+\frac{i}{6} \left[ \nabla_{\dot A} \left( \nabla^2 + \frac{1}{3} R \right) 
U \right] \varepsilon_A^{\ \ B} \nonumber \\
&+&\frac{i}{4} \varepsilon_A^{\ \ B} \nabla_{\dot A} \left( \nabla^{D \dot D}
\chi_{D \dot D} \right) 
+\frac{1}{24} \varepsilon_A^{\ \ B} \nabla_{\dot A} \left( \nabla_D 
\nabla_{\dot D} -\nabla_{\dot D} \nabla_D \right) \chi^{D \dot D} \nonumber \\
&+&\frac{1}{8} \nabla_{\dot A} \left( \nabla_{\dot D} \nabla_A
-\nabla_A \nabla_{\dot D} \right) \chi^{B \dot D} 
-\frac{1}{16} \overline{R} \nabla_A \chi_{\ \ \dot A}^B 
-\frac{i}{4} \nabla_{A \dot B} \nabla_{\dot A} \chi^{B \dot B} \nonumber \\
&-&\frac{1}{24} \chi_{\ \ \dot A}^B \nabla_A \overline{R} 
+\frac{3}{16} G_A^{\ \ \dot B} \nabla_{\dot B} \chi_{\ \ \dot A}^B 
+\frac{1}{16} G_{A \dot A} \nabla_{\dot B} \chi^{B \dot B} \nonumber \\
&-&\frac{1}{16} G_{A \dot B} \nabla_{\dot A} \chi^{B \dot B} 
-\frac{3}{8} G^{B \dot B} \nabla_{\dot A} \chi_{A \dot B} 
-\frac{3}{8} \varepsilon_A^{\ \ B} \left( \nabla_{\dot A} \chi^{C \dot B} 
\right) G_{C \dot B}
\end{eqnarray}

This finishes our computation of $H_M^{\ \ N}$. We see that we did not need to
introduce any other arbitrary superfield for this computation: the superfields
introduced in section \ref{2} are enough, as expected, because we have three 
independent superfields and we want to find out relations between them. These 
relations should be generalizations of (\ref{g=0}) and (\ref{r=0}):
$G_{A \dot A}$ and $R$ should be functions of the independent superfield
$W_{ABC}$, which is responsible to the correction terms in (\ref{action}).
Therefore we only have two equations and two arbitrary superfields. 

\subsection{The field equations in superspace}
\label{42}
\indent
The variation of our action (\ref{action}) is given by the superspace integral
of
\begin{eqnarray}
\delta \left[ E\left( 1+\alpha W^2\overline{W}^2\right) \right] &=&
2\alpha E\left(\overline{W}^2W^{ABC}\delta W_{ABC}+W^2 
W^{\dot A\dot B\dot C}\delta W_{\dot A\dot B\dot C}\right) \nonumber \\
&+& \left( 1+\alpha W^2\overline{W}^2\right) \delta E \label{deltaaction}
\end{eqnarray}

For this computation we obviously need the constrained
variation of $W_{ABC}$. The details of this calculation are presented in appendix 
\ref{b}, where we derive an equation for $\overline{W}^2W^{ABC}\delta W_{ABC}
+ \mathrm{h.c.}$. Having this result (which is 
eq. (\ref{2wdw})), we should now integrate it in superspace (multiplied by 
$E$). In order to "factorize" $i \chi^{A\dot A}, U, \overline{U}$ to get the 
field equations, we should integrate by parts the terms which contain 
derivatives of these superfields. Following this procedure, we write 
\begin{eqnarray}
&\left. \right.&
\int E \left[ 2\overline{W}^2 W^{ABC} \delta W_{ABC} +2W^2 W^{\dot A\dot B\dot
C}\delta W_{\dot A\dot B\dot C} \right] d^4x d^4\theta \nonumber \\
&=& \int E \left[ i\chi^{A\dot A}\left[ -2G_{A\dot A}W^2\overline{W}^2 
+\frac{i}{8} \nabla^B\left( \overline{R}\nabla_{\ \ \dot A}^C
\left( W_{ABC}\overline{W}^2\right) \right) 
\right. \right. \nonumber \\
&+&\frac{1}{2}\left( \nabla_{\dot D}\nabla^B-\nabla^B\nabla_{\dot D}\right)
\left( \left( \nabla_{\dot A}G^{C \dot D}\right)
W_{ABC}\overline{W}^2\right) \nonumber \\
&-&\frac{i}{4} \left( \nabla_{\dot A}\nabla^B-\nabla^B\nabla_{\dot
A}\right) \nabla^{\dot D}\nabla_{\ \ \dot D}^C\left( W_{ABC}\overline{W}^2
\right) \nonumber \\
&+&\frac{1}{2}\nabla^{\dot D}\nabla_{\ \ \dot A}^B \nabla_{\ \ \dot D}^C\left(
W_{ABC}\overline{W}^2\right) 
-\frac{i}{12}\left( \nabla^B\overline{R}\right)
\nabla_{\ \ \dot A}^C\left( W_{ABC}\overline{W}^2\right) \nonumber \\
&+&\frac{3}{8}i\nabla^{\dot D}\left( G_{\ \ \dot D}^B\nabla_{\ \ \dot
A}^C \left(W_{ABC}\overline{W}^2\right) \right) 
-\frac{i}{8} \nabla_{\dot A}\left( G^{B\dot D}\nabla_{\ \ \dot D}^C\left(
W_{ABC}\overline{W}^2\right) \right) \nonumber \\ 
&-&\frac{5}{8} i\nabla^{\dot D}\left( G_{\ \ \dot A}^B\nabla_{\ \ \dot D}^C
\left(W_{ABC}\overline{W}^2\right) \right) 
+ \frac{1}{12}\left( \nabla^B G_{\ \ \dot A}^C\right)
\overline{R}W_{ABC}\overline{W}^2 \nonumber \\
&-&\frac{3}{16}\nabla^B\left( G_{\ \ \dot
A}^C\overline{R}W_{ABC}\overline{W}^2\right)
+ \frac{1}{8}\left( \nabla^B
\overline{R}\right) G_{\ \ \dot A}^C W_{ABC}\overline{W}^2 \nonumber \\
&+& \frac{3}{8} \left( \nabla_{\dot A}\nabla^B-\nabla^B\nabla_{\dot A}\right) 
\nabla^{\dot D}\left( G_{\ \ \dot D}^C W_{ABC} \overline{W}^2\right) 
\nonumber \\
&+& \frac{3}{4} i \nabla^{\dot D} \nabla_{\ \ \dot A}^C \left(G_{\ \ \dot
D}^B W_{ABC}\overline{W}^2\right) 
+\left. \frac{3}{8} \nabla^{\dot D} \left( G_{\ \ \dot
A}^B G_{\ \ \dot D}^C W_{ABC} \overline{W}^2\right) + \mathrm{h.c.} \right]
\nonumber \\
&-& U \nabla^2\left( W^2\overline{W}^2\right) 
-\frac{U}{3} R W^2\overline{W}^2 
- \overline{U}\overline{\nabla }^2\left(W^2\overline{W}^2\right) 
- \frac{\overline{U}}{3} \overline{R}W^2\overline{W}^2
\nonumber \\
&+& \left.\mbox{full divergences} \right] 
d^4x d^4\theta \label{2ewdw}
\end{eqnarray}

Using the previous results (\ref{deltapure}), (\ref{deltapure3})
and (\ref{2ewdw}), and remembering that we can use (\ref{div}) to discard the terms
which are full divergences, we finally get for (\ref{deltaaction})

\begin{eqnarray}
\int \delta \left[ E\left( 1+\alpha W^2\overline{W}^2\right) \right]
d^4x d^4\theta &=& \alpha \int E \left\{
\left[ \frac{2}{3\alpha }G_{A\dot A}-\frac{4}{3}G_{A\dot
A}W^2\overline{W}^2 \right. \right. \nonumber \\
&-&\frac{1}{12}\left( \nabla_A \nabla_{\dot A}-\nabla_{\dot A}\nabla_A \right) 
\left( W^2\overline{W}^2\right)  \nonumber \\ 
&+&\left[ \frac{1}{2} \left( \nabla_{\dot D} \nabla^B- \nabla^B \nabla_{\dot
D}\right) \left( \left( \nabla_{\dot A} G^{C \dot D}\right)
W_{ABC}\overline{W}^2\right) \right. \nonumber \\
&+& \frac{i}{8}\nabla^B\left(\overline{R}\nabla_{\ \ \dot A}^C
\left(W_{ABC}\overline{W}^2 \right) \right) \nonumber \\
&-&\frac{i}{4} \left( \nabla_{\dot A}\nabla^B-\nabla^B\nabla_{\dot A}\right) 
\nabla^{\dot D}
\nabla_{\ \ \dot D}^C \left(W_{ABC} \overline{W}^2 \right) \nonumber \\
&+&\frac{1}{2} \nabla^{\dot D} \nabla_{\ \ \dot A}^B 
\nabla_{\ \ \dot D}^C \left(W_{ABC} \overline{W}^2\right) \nonumber \\
&-& \frac{i}{12} \left(\nabla^B \overline{R} \right)
\nabla_{\ \ \dot A}^C \left(W_{ABC} \overline{W}^2\right) \nonumber \\
&+&\frac{3}{8}i \nabla^{\dot D} \left(G_{\ \ \dot D}^B 
\nabla_{\ \ \dot A}^C \left(W_{ABC} \overline{W}^2 \right) \right) \nonumber \\
&-&\frac{i}{8} \nabla_{\dot A} \left(G^{B \dot D} 
\nabla_{\ \ \dot D}^C \left(W_{ABC} \overline{W}^2 \right) \right) \nonumber \\
&-& \frac{5}{8} i \nabla^{\dot D} \left(G_{\ \ \dot A}^B \nabla_{\ \ \dot D}^C
\left(W_{ABC} \overline{W}^2\right) \right) \nonumber \\
&-& \frac{3}{16} \nabla^B \left(G_{\ \ \dot A}^C \overline{R} W_{ABC} 
\overline{W}^2\right) \nonumber \\ 
&+& \frac{1}{12} \left(\nabla^B G_{\ \ \dot A}^C \right) \overline{R} 
W_{ABC} \overline{W}^2 \nonumber \\
&+& \frac{1}{8}\left(\nabla^B \overline{R} \right) G_{\ \ \dot A}^C
W_{ABC}\overline{W}^2 \nonumber \\
&+&\frac{3}{8} \left(\nabla_{\dot A} \nabla^B -\nabla^B
\nabla_{\dot A} \right) \nabla^{\dot D} \left(G_{\ \ \dot D}^C W_{ABC}
\overline{W}^2 \right) \nonumber \\
&+& \frac{3}{4}i \nabla^{\dot D} \nabla_{\ \ \dot A}^C
\left( G_{\ \ \dot D}^B W_{ABC} \overline{W}^2 \right) \nonumber \\ 
&+& \left. \left. \frac{3}{8} \nabla^{\dot D} \left(G_{\ \ \dot A}^B 
G_{\ \ \dot D}^C W_{ABC} \overline{W}^2 \right) + {\mathrm h.c.} \right] 
\right] i\chi^{A\dot A} \nonumber \\
&-&\frac{2}{3} U\overline{W}^2 \nabla^2 W^2 -\frac{2}{9} U R W^2
\overline{W}^2 
+ \frac{1}{9} \frac{UR}{\alpha} \nonumber \\
&-& \left. \frac{2}{3}\overline{U} W^2
\overline{\nabla}^2 \overline{W}^2 
-\frac{2}{9} \overline{U} \overline{R}
W^2 \overline{W}^2 
+\frac{1}{9} \frac{\overline{U}\overline{R}}{\alpha}
\right\} d^4x d^4\theta 
\nonumber \\ &=&0
\end{eqnarray}

The $U, \overline{U}$ field equations are immediately read:

\begin{eqnarray}
\frac{2}{3}U\overline{W}^2\nabla^2 W^2 +\frac{2}{9} UR W^2 \overline{W}^2
-\frac{1}{9} \frac{UR}{\alpha}=0
\nonumber \\
\frac{R}{3} \left( 2W^2\overline{W}^2 - \frac{1}{\alpha }\right)
=-2\overline{W}^2 \nabla^2 W^2
\end{eqnarray}

Since $W^5=0$, we get then the following (exact) result:

\begin{eqnarray}
R&=&6\alpha \frac{\overline{W}^2 \nabla^2 W^2}{1-2\alpha W^2
\overline{W}^2}=6\alpha \overline{W}^2 \nabla^2 W^2+12\alpha^2
\overline{W}^4 W^2 \nabla^2 W^2 \nonumber \\
\overline{R}&=&6\alpha \frac{W^2 \overline{\nabla}^2
\overline{W}^2}{1-2\alpha
W^2 \overline{W}^2}=6\alpha W^2 \overline{\nabla }^2
\overline{W}^2+12\alpha
^2 W^4 \overline{W}^2 \overline{\nabla }^2 \overline{W}^2 \label{r}
\end{eqnarray}

The $\chi^{A\dot A}$ field equation can also be immediately read:

\begin{eqnarray}
\frac{G_{A\dot A}}{\alpha }&=&
2G_{A\dot A} W^2 \overline{W}^2 
+\frac{i}{4}
\overline{W}^2 \nabla_{A \dot A}W^2
-\frac{i}{4} W^2 \nabla_{A \dot A}
\overline{W}^2 
+\frac{1}{4} \nabla_A \left(W^2 \right) \nabla_{\dot A}
\overline{W}^2 \nonumber \\
&-&\left[- \frac{i}{8} \left( \nabla^B \overline{R} \right)
\nabla_{\ \ \dot A}^C \left(W_{ABC} \overline{W}^2\right) 
+\frac{3}{16} \left(\nabla^B \overline{R} \right) G_{\ \ \dot A}^C W_{ABC} 
\overline{W}^2  \right. \nonumber \\
&-&\frac{9}{32} \nabla^B \left(\overline{R} G_{\ \ \dot A}^C W_{ABC}
\overline{W}^2 \right) 
+\frac{9}{16} \nabla^{\dot D} \left( G_{\ \ \dot A}^B
G_{\ \ \dot D}^C W_{ABC} \overline{W}^2 \right) \nonumber \\
&+& \frac{9}{16} \left(\nabla_{\dot A} \nabla^B-\nabla^B \nabla_{\dot A}
\right) \nabla^{\dot D} \left(G_{\ \ \dot D}^C W_{ABC} \overline{W}^2 \right) 
\nonumber \\
&-&\frac{3}{4} \left(\nabla_{\dot D} \nabla^B-\nabla^B \nabla_{\dot
D} \right) \left( \left( \nabla_{\dot A} G^{C \dot D} \right) W_{ABC}
\overline{W}^2\right) \nonumber \\
&+& \frac{9}{8}i \nabla^{\dot D} \nabla_{\ \ \dot A}^B \left( G_{\ \ \dot D}^C 
W_{ABC} \overline{W}^2\right) 
+\frac{3}{16} i \nabla^B \left( \overline{R}
\nabla_{\ \ \dot A}^C \left(W_{ABC} \overline{W}^2\right) \right) \nonumber \\
&+&\frac{1}{8} \overline{R} \left( \nabla^B G_{\ \ \dot A}^C \right)
W_{ABC} \overline{W}^2 
+ \frac{9}{16} i \nabla^{\dot D} \left(G_{\ \ \dot
D}^C \nabla_{\ \ \dot A}^B \left(W_{ABC}\overline{W}^2 \right) \right)
\nonumber \\
&-&\frac{3}{16} i \nabla_{\dot A} \left(G^{B\dot D} \nabla_{\ \ \dot D}^C
\left(W_{ABC}\overline{W}^2\right) \right) 
+ \frac{3}{4}\nabla^{\dot D} \nabla_{\ \ \dot A}^B\nabla_{\
\ \dot D}^C \left(W_{ABC} \overline{W}^2\right) \nonumber \\
&-&\frac{15}{16}i \nabla^{\dot D} \left( G_{\ \ \dot A}^B \nabla_{\ \ \dot D}^C \left( W_{ABC} \overline{W}^2 \right) \right) \nonumber \\
&-& \left.
\frac{3}{8} i \left( \nabla_{\dot A} \nabla^B - \nabla^B \nabla_{\dot A}
\right) \nabla^{\dot D}
\nabla_{\ \ \dot D}^C \left(W_{ABC} \overline{W}^2\right) 
+{\mathrm h.c.} \right] \label{gcomplicated}
\end{eqnarray}
This equation must be rewritten in a different form. Using the 
supercommutation relations and the solutions for torsions and curvatures in 
appendix \ref{appendix1.2}, most of its terms may be rewritten in such
a way that they contain a minimal number of derivatives and an explicit 
dependence on $R, G_{A \dot A}, W_{ABC}$. The actual computations for this 
purpose are very heavy, and nothing special can
be learnt from them. We present the intermediate results (the expansion of
each term in (\ref{gcomplicated})) in appendix \ref{c}.
Here we present only the final results. Replacing each expression of appendix 
\ref{c} in (\ref{gcomplicated}), we obtain the following expanded field 
equation for $G_{A\dot A}$:

\begin{eqnarray}
G_{A \dot A} &=&
\frac{1}{4} \frac{\alpha}{1+ \alpha W^2 \overline{W}^2}
\left( \nabla_A W^2 \right) \nabla_{\dot A} \overline{W}^2 
\nonumber \\
&-&\frac{\alpha}{1+ \alpha W^2 \overline{W}^2}
\left[i W^2 \nabla_{A \dot A} \overline{W}^2 
+ \frac{3}{2} \left( \nabla^{\dot D} \nabla_{\ \ \underline{\dot A}}^B
\nabla_{\ \ \underline{\dot D}}^C \overline{W}^2 \right) W_{ABC}
\right. \nonumber \\
&+&\frac{3}{8} W_A^{\ \ BC} \left( \nabla^D W_{DBC} \right)
\nabla_{\dot A} \overline{W}^2  
\nonumber \\
&-&\frac{i}{8} \overline{W}^2 \left( \nabla^B \nabla_{\ \ \dot A}^C
\overline{R} \right) W_{ABC} 
-\frac{i}{8} \overline{W}^2 \left( \nabla_{\ \ \dot A}^C \overline{R} 
\right) \nabla^B W_{ABC} \nonumber \\
&-& \frac{i}{8} \left( \nabla^B \overline{R} \right) \overline{W}^2 
\nabla_{\ \ \dot A}^C W_{ABC}
-\frac{9}{32}i \left( \nabla^B \overline{R} \right) \left( \nabla_{\ \ \dot
A}^C \overline{W}^2 \right) W_{ABC} \nonumber \\
&-& \frac{1}{16} \overline{W}^2 G_{\ \ \dot A}^C \left( \nabla^B \overline{R} 
\right) W_{ABC}
+ \frac{1}{32} \left( \nabla^C \overline{R} \right) \left( \nabla_{\dot A} 
\overline{W}^2 \right) \nabla^B W_{ABC} \nonumber \\
&-&\frac{3}{16} i \overline{R} \left( \nabla_{\ \ \dot A}^C \overline{W}^2 
\right) \nabla^B W_{ABC} 
-\frac{1}{16} \overline{R} \overline{W}^2 \left( \nabla^B G_{\ \ \dot A}^C 
\right) W_{ABC} \nonumber \\
&-&\frac{i}{8} \overline{R} \overline{W}^2 \nabla^B \nabla_{\ \ \dot A}^C
W_{ABC}
- \frac{1}{16} \overline{R} \overline{W}^2 G_{\ \ \dot A}^C \nabla^B
W_{ABC} \nonumber \\
&+& \frac{3}{2} i\left( \nabla_{\ \ \underline{\dot A}}^B G_{\ \ 
\underline{\dot D}}^C \right) \left( \nabla^{\dot D} \overline{W}^2 \right)
W_{ABC} \nonumber \\
&-& \frac{3}{4} \left( \nabla^B \nabla_{\underline{\dot A}} 
G_{\ \ \underline{\dot D}}^C \right) \left( \nabla^{\dot D} \overline{W}^2
\right) W_{ABC} \nonumber \\
&-& \frac{9}{2}i \left( \nabla^{B\dot D} \overline{W}^2
\right) \left( \nabla_{\underline{\dot A}} G_{\ \ \underline{\dot D}}^C
\right) W_{ABC} \nonumber \\
&+& \frac{3}{2} \left( \nabla^{\dot D} \overline{W}^2 \right)
\left( \nabla_{\underline{\dot A}} G_{\ \ \underline{\dot D}}^C \right)
\nabla^B W_{ABC} \nonumber \\
&+& \frac{3}{16} \left( \overline{\nabla}^2 \overline{W}^2
\right) \left( \nabla ^B G_{\ \ \dot A}^C \right) W_{ABC}
-\frac{9}{2} G_{\ \
\underline{\dot A}}^B G_{\ \ \underline{\dot D}}^C \left( \nabla^{\dot D}
\overline{W}^2 \right) W_{ABC} \nonumber \\
&-& 3i G^{B \dot D} \left(
\nabla_{\underline{\dot A}} \overline{W}^2 \right) \nabla_{\ \
\underline{\dot D}}^C W_{ABC}
+ \frac{3}{2}i G_{\ \ \dot A}^B \left( \nabla^{\dot D} \overline{W}^2 \right) 
\nabla_{\ \ \dot D}^C W_{ABC} \nonumber \\
&-&\frac{9}{2} iG^{B\dot D} \left( \nabla_{\underline{\dot A}} \nabla_{\ \
\underline{\dot D}}^C \overline{W}^2 \right) W_{ABC} 
+3i G_{\ \ \dot A}^B \left( \nabla^{\dot D} \nabla_{\ \ \dot D}^C 
\overline{W}^2 \right) W_{ABC} \nonumber \\
&+& \frac{3}{2} \left( \nabla^{\dot D} \nabla_{\ \ \dot D}^C \overline{W}^2
\right) \nabla_{\ \ \dot A}^BW_{ABC} 
-\frac{3}{8}i \left( \overline{\nabla}^2 \overline{W}^2 \right) \nabla^B 
\nabla_{\ \ \dot A}^C W_{ABC} \nonumber \\
&-& \left. \frac{3}{8}i \left(\overline{\nabla }^2 \nabla_{\ \ \dot A}^B
\overline{W}^2 \right) \nabla^C W_{ABC} +\mathrm{h.c.} \right] 
\label{gexpanded}
\end{eqnarray}
This is a complicated nonlinear differential equation for $G_{A \dot A}$, 
which cannot be solved in a closed form. The best we can do is to solve it as 
a power series in $\alpha$. The main question is: does that expansion stop for 
some power of $\alpha$, like in $R$, or is it an infinite expansion?

To answer this question, we must look at the different terms of (\ref{gexpanded}). Expanding the overall factor
\begin{equation}
\frac{\alpha}{1+ \alpha W^2 \overline{W}^2} = \alpha - \alpha^2 W^2 
\overline{W}^2 + \alpha^3 W^4 \overline{W}^4 \label{factor} 
\end{equation}
and keeping in mind that $W^5=0$, we see that it is enough to multiply 
(\ref{factor}) by a term with a free $W_{ABC}$ factor to be sure that the expansion stops at 
$\alpha^2$, like $R$. Therefore, if each and every term of (\ref{gexpanded}) 
has, after each iteration, a free $W_{ABC}$ factor (or something from which it can be extracted, like 
{\em one} derivative of $W^2$), we have a finite polynomial expansion of 
$G_{A\dot A}$. If at least one of the terms of (\ref{gexpanded}) does not have 
such a factor, the iteration process will never stop and we will have an 
infinite expansion.

The problems with getting a free $W_{ABC}$ factor after iteration arise from the terms with derivatives of $G_{A\dot A}$, because these terms may contribute with derivatives of $W_{ABC}$, which do not stop the iteration process.

Let's then start by analyzing each term of (\ref{gexpanded}) which does {\it not} include derivatives of $G_{A\dot A}$. By inspection, we conclude that all of them have a $W_{ABC}$ factor with the exception of the three last terms which are only 
proportional to $\nabla^C W_{ABC}$ or its derivatives. Therefore, using 
(\ref{diffw}), we can write them as derivatives of $G_{A \dot A}$ plus terms 
proportional to $W_{ABC}$:
\begin{equation}
\nabla_{\ \ \dot A}^B W_{ABC}=\frac{i}{2} \nabla_{\dot A} \nabla^B W_{ABC}
+\frac{5}{2}i G_{\ \ \dot A}^B W_{ABC} = \nabla_{\dot A}
\nabla^{\ \ \dot B}_{\underline{A}} G_{\underline{C} \dot B}
+\frac{5}{2}i G_{\ \ \dot A}^B W_{ABC} \label{dbdaw}
\end{equation}

\begin{equation}
\nabla^B \nabla_{\ \ \dot A}^C W_{ABC}= -2i \nabla^B_{\ \ \dot A}
\nabla^{\ \ \dot B}_{\underline{A}} G_{\underline{B} \dot B} +
G^B_{\ \ \dot A} \nabla^{\ \ \dot B}_{\underline{A}} G_{\underline{B} \dot B}
+\frac{11}{2}i \left(\nabla^B G_{\ \ \dot A}^C \right) W_{ABC} \label{dbcdaw}
\end{equation}
We have two kinds of terms to iterate in (\ref{gexpanded}): the ones without and the ones with derivatives of $G_{A\dot A}$. Replacing (\ref{dbdaw}) and (\ref{dbcdaw}) in (\ref{gexpanded}), we conclude that the former terms all have a $W_{ABC}$ factor after iteration and only have a finite contribution to the $\alpha$ expansion of $G_{A\dot A}$. 

The terms of (\ref{gexpanded}) with derivatives of $G_{A\dot A}$ are linearly independent and cannot be simplified. Each time we iterate a solution for $G_{A\dot A}$ of a certain order in $\alpha$ on each of these terms, we get terms with higher derivatives of $W_{ABC}$ and no new factors of $W_{ABC}$
itself. This means that, because of these terms, the actual solution for $G_{A \dot A}$ is, as 
opposite to $R$, an infinite series in $\alpha$, with derivatives of 
$W_{ABC}$ to all orders.

From what was mentioned in section \ref{3}, this result was expected. The
nonalgebraic field equation for the auxiliary field $A_m$, because of
its higher-derivative terms in the component action, is actually obtained by 
taking the $\theta=0$ component of (\ref{gexpanded}).


\subsection{Computation of $G_{A\dot A}$ to second order in $\alpha$}
\indent

From the results of the previous subsection, we conclude that the complete 
on-shell
supersymmetrization of ${\cal R}^4$ requires an infinite number of terms.
In practice, what we do is to solve (\ref{gexpanded}) for $G_{A\dot A}$
perturbatively order by order in $\alpha$, by iterating, for $n$-th order, the 
$n-1$-th solution.

In this subsection, we solve (\ref{gexpanded}) for $G_{A\dot A}$ up to second 
order in $\alpha$ (the order at which $R$ stops). We must then first solve for 
$G_{A\dot A}$ to first order in $\alpha$.

In order to do that first we take, in (\ref{gexpanded}), only the terms which 
are of first order in $\alpha$. We recall that, to order 0 in $\alpha$ (pure 
supergravity), both $G_{A \dot A}=0$ and $R=0$. The first non-trivial order 
is, then, $\alpha$, which means that both $G_{A \dot A}$ and $R$ necessarily 
contribute with, at least, one power of $\alpha$. We also recall that, due to 
the off-shell identity (\ref{diffw}), terms like $\nabla^C W_{ABC}$ are, at 
least, of order $\alpha$, and therefore
\begin{equation}
\nabla_A W_{BCD}= \nabla_{\underline{A}} W_{\underline{B} \underline{C} 
\underline{D}} + {\cal O}\left( \alpha\right)
\end{equation}
We then get

\begin{eqnarray}
G_{A\dot A}&=& 
-i\alpha W^2 \nabla_{A\dot A} \overline{W}^2 
+i \alpha \overline{W}^2 \nabla_{A\dot A} W^2
+ \frac{\alpha}{4} \left(\nabla_A W^2 \right) 
\nabla_{\dot A} \overline{W}^2  \nonumber \\
&-&\frac{3}{2} \alpha \left( \nabla^B \nabla_{\underline{A}}^{\ \ \dot B}
\nabla_{\underline{B}}^{\ \ \dot C} W^2\right) W_{\dot A\dot B\dot C} 
\nonumber \\
&-&\frac{3}{2} \alpha \left( \nabla^{\dot B} \nabla_{\ \ \underline{\dot A}}^B
\nabla_{\ \ \underline{\dot B}}^C \overline{W}^2 \right) W_{ABC}
+{\cal O}\left(\alpha^2 \right) \label{g1}
\end{eqnarray}
which satisfies
\begin{equation}
\nabla^{\dot A} G_{A\dot A} =-\frac{1}{4} \alpha \left( \nabla_A W^2
\right)
\overline{\nabla }^2 \overline{W}^2 -i \alpha W^2 \nabla^{\dot A} \nabla_{A
\dot A} \overline{W}^2 +{\cal O}\left( \alpha ^2\right)
\end{equation}

We now do the same with (\ref{r}):
\begin{equation}
\overline{R}=6 \alpha W^2 \overline{\nabla }^2 \overline{W}^2 +{\cal O}\left(
\alpha^2\right) \label{r1} 
\end{equation}
By taking
\begin{equation}
\nabla_A \overline{R}= 6 \alpha \left( \nabla_A W^2 \right)
\overline{\nabla}^2 \overline{W}^2 +24 i \alpha W^2 \nabla^{\dot A}
\nabla_{A \dot A} \overline{W}^2+{\cal O}\left( \alpha^2\right) 
\end{equation}
we see that the off-shell relation (\ref{diffg}) is satisfied to first order 
in $\alpha$, as it should. 

We now proceed keeping, from (\ref{gexpanded}), only the terms which are of 
order $\alpha^2$. These terms are:

\begin{eqnarray}
G_{A\dot A}&=& - \alpha  W^2 \overline{W}^2 G_{A\dot A} + \frac{\alpha}{4} 
\left(\nabla_A W^2 \right) \nabla_{\dot A} \overline{W}^2  \nonumber \\
&+& \left[i \alpha \overline{W}^2 \nabla_{A\dot A} W^2 - \frac{3}{2} \alpha 
\left( \nabla^{\dot B} \nabla_{\ \ \underline{\dot A}}^B
\nabla_{\ \ \underline{\dot B}}^C \overline{W}^2 \right) W_{ABC} \right. 
\nonumber \\
&-& \frac{3}{8} \alpha \left(\nabla^D W_D^{\ \ BC} \right) W_{ABC} 
\nabla_{\dot A} \overline{W}^2 
+ \frac{i}{8} \alpha \overline{W}^2 \left( \nabla^B \nabla_{\ \ \dot A}^C 
\overline{R} \right) W_{ABC}
 \nonumber \\
&+& \frac{9}{32}i \alpha \left( \nabla^B \overline{R} \right) 
\left( \nabla_{\ \ \dot A}^C \overline{W}^2 \right) W_{ABC}
-\frac{3}{2} i\alpha \left( \nabla_{\ \ \underline{\dot A}}^B G_{\ \ 
\underline{\dot D}}^C \right) \left( \nabla^{\dot D} \overline{W}^2 \right)
W_{ABC} \nonumber \\
&+& \frac{3}{4} \alpha \left( \nabla^B \nabla_{\underline{\dot A}} 
G_{\ \ \underline{\dot D}}^C \right) \left( \nabla^{\dot D} \overline{W}^2
\right) W_{ABC} \nonumber \\
&-& \frac{3}{16} \alpha \left( \overline{\nabla}^2 \overline{W}^2
\right) \left( \nabla^B G_{\ \ \dot A}^C \right) W_{ABC} 
+ \frac{9}{2} i \alpha G^{B\dot D} \left( \nabla_{\underline{\dot A}} 
\nabla_{\ \ \underline{\dot D}}^C \overline{W}^2 \right) W_{ABC} \nonumber \\
&-&3i \alpha G_{\ \ \dot A}^B \left( \nabla^{\dot D} \nabla_{\ \ \dot D}^C 
\overline{W}^2 \right) W_{ABC} 
- \frac{3}{2} \alpha \left( \nabla^{\dot D} \nabla_{\ \ \dot D}^C 
\overline{W}^2 \right) \nabla_{\ \ \dot A}^BW_{ABC} \nonumber \\
&+& \frac{9}{2}i \alpha \left( \nabla^{B\dot D} \overline{W}^2
\right) \left( \nabla_{\underline{\dot A}} G_{\ \ \underline{\dot D}}^C
\right) W_{ABC} +\frac{3}{8}i \alpha \left( \overline{\nabla}^2 \overline{W}^2 
\right) \nabla^B \nabla_{\ \ \dot A}^C W_{ABC} \nonumber \\
&+& \left. \frac{3}{8}i \alpha \left(\overline{\nabla }^2 \nabla_{\ \ \dot A}^B
\overline{W}^2 \right) \nabla^C W_{ABC} +\mathrm{h.c.} \right] 
+{\cal O}\left( \alpha ^3\right) \label{g2a}
\end{eqnarray}
We must now replace the solutions (\ref{g1}) and (\ref{r1})
for $G_{A\dot A}$ and $R$ in (\ref{g2a}). For that, we need a series of 
intermediate expressions which we present in appendix \ref{d}.
The actual computations are very heavy, and nothing special can
be learnt from them. 
Here we present only the final results. Replacing each expression of appendix 
\ref{d} in (\ref{g2a}), we obtain the following expanded field 
equation for $G_{A\dot A}$ \footnote{When we double-underline a pair of 
indices, like $B, C$ in $W^{\underline{\underline{B}}\underline{D}E} 
W^{\underline{\underline{C}}}_{\ \ E\underline{F}}$, we mean that we are 
symmetrizing that pair of indices and there is another pair of indices of the 
same type ($D, F$ in our case) which is being symmetrized independently of the
first.}:

\begin{eqnarray}
G_{A\dot A}&=& \frac{1}{4} \alpha \left(\nabla_A W^2 \right) 
\nabla_{\dot A} \overline{W}^2  - \frac{1}{4} \alpha^2 W^2 \overline{W}^2
\left(\nabla_A W^2 \right) \nabla_{\dot A} \overline{W}^2 \nonumber \\
&+& \left[i \alpha \overline{W}^2 \nabla_{A\dot A} W^2 - \frac{3}{2} \alpha 
\left( \nabla^{\dot B} \nabla_{\ \ \dot B}^B
\nabla_{\ \ \dot A}^C \overline{W}^2 \right) W_{ABC} \right. 
\nonumber \\
&-& \frac{3}{8} \alpha \left(\nabla^D W_D^{\ \ BC} \right) W_{ABC} 
\nabla_{\dot A} \overline{W}^2 -\frac{3}{2} \alpha 
\left( \nabla^{\dot D} \nabla_{\ \ \dot D}^C 
\overline{W}^2 \right) \nabla_{\ \ \dot A}^BW_{ABC}  \nonumber \\
&+&  \frac{3}{8}i \alpha \left( \overline{\nabla}^2 \overline{W}^2 
\right) \nabla^B \nabla_{\ \ \dot A}^C W_{ABC} 
+ \frac{3}{8}i \alpha \left(\overline{\nabla }^2 \nabla_{\ \ \dot A}^B
\overline{W}^2 \right) \nabla^C W_{ABC} \nonumber \\
&-&2i \alpha^2 W^4 \overline{W}^2 \nabla_{A\dot A} W^2 - \frac{3}{2} 
\alpha^2 W^2 \overline{W}^2
\left( \nabla^{\dot B} \nabla_{\ \ \dot B}^B
\nabla_{\ \ \dot A}^C \overline{W}^2 \right) W_{ABC} \nonumber \\
&+& 9 \alpha^2 W^2 \left(\nabla^{B \dot B} \overline{W}^2 \right) \left( 
\nabla_{\underline{\dot A}} \nabla_{\ \ \underline{\dot B}}^C 
\overline{W}^2 \right) W_{ABC} \nonumber \\
&-& \frac{39}{4} \alpha^2 W^2 \left(
\nabla_{\ \ \dot A}^B \overline{W}^2 \right) \left( \nabla^{\dot D} 
\nabla_{\ \ \dot D}^C \overline{W}^2 \right) W_{ABC} \nonumber \\
&-& 3 \alpha^2 W^2 \left(\nabla_{\ \ \underline{\dot A}}^B
\nabla_{\ \ \underline{\dot B}}^C \overline{W}^2 \right) \left(\nabla^{\dot B} 
\overline{W}^2 \right) W_{ABC} \nonumber \\
&+& 18 \alpha^2 W^{BDE} W^C_{\ \ EF} W_{ABC} \left(\nabla_{\underline{D} 
\underline{\dot A}} \nabla_{\ \ \underline{\dot B}}^{\underline{F}}
\overline{W}^2 \right) \nabla^{\dot B}\overline{W}^2 \nonumber \\
&-& \frac{9}{2} i \alpha^2 \left(\nabla^{\dot C} 
\nabla_{\ \ \dot C}^{\underline{B}}
\nabla_{\ \ \underline{\dot A}}^{\underline{E}}
\nabla_{\ \ \underline{\dot B}}^{\underline{F}} \overline{W}^2 \right)
W^C_{\ \ EF} W_{ABC} \nabla^{\dot B} \overline{W}^2 \nonumber \\
&-& \frac{9}{32} i \alpha^2 \left(\overline{\nabla}^2 \overline{W}^2 \right)
\left(\nabla_{A \dot A} \overline{W}^2 \right) W^{FDE} W_F^{\ \ BC} 
\nabla_{\underline{B}} W_{\underline{C} \underline{D} \underline{E}} 
\nonumber \\
&+& \frac{27}{8} i \alpha^2 W^{EFC} W_{ABC} \left(\nabla^{B \dot B} 
\overline{W}^2 \right) \overline{\nabla}^2 \nabla_{E \underline{\dot A}} 
\nabla_{F \underline{\dot B}} \overline{W}^2 
\nonumber \\
&-& \frac{27}{4} i \alpha^2 W^{BDE} W_{ABC} \left(\nabla^{\dot C} 
\nabla_{D \dot C} \nabla_{E \dot B} \overline{W}^2 \right) 
\nabla_{\underline{\dot A}} \nabla^{C \underline{\dot B}} \overline{W}^2 
\nonumber \\
&-& \frac{9}{2} i \alpha^2 W^{BDE} W_{ABC} \left(\nabla^{\dot B} 
\nabla_{D \dot B} \nabla_{E \dot A} \overline{W}^2 \right)
\nabla^{\dot C} \nabla^C_{\ \ \dot C} \overline{W}^2 \nonumber \\
&-& 9 \alpha^2 \left(\nabla^{\dot D} \overline{W}^2 \right) 
W_{\ \ \underline{\dot A} \underline{\underline{\dot B}}}^{\dot E} W_{\dot E 
\underline{\underline{\dot C}} \underline{\dot D} }
W_{ABC} \nabla^{B \dot B} \nabla^{C \dot C} W^2 
\nonumber \\
&+& \frac{27}{2} \alpha^2 \left(\nabla^{C \dot C} \overline{W}^2 \right) 
W_{\dot A \dot B \dot C} W^{BEF} W_{ABC} \nabla_E \nabla_F^{\ \ \dot B} W^2
\nonumber \\
&+& \frac{3}{2} \alpha^2 \overline{W}^2 
\left( \nabla^{\dot D} \overline{W}^2 \right) W_{ABC}
\nabla^B_{\ \ \underline{\dot A}} \nabla^C_{\ \ \underline{\dot B}} W^2 
\nonumber \\
&-& \frac{9}{2} \alpha^2 \overline{W}^2 
\left( \nabla_{\underline{\dot A}} 
\nabla^B_{\ \ \underline{\dot B}}\overline{W}^2 \right)
W_{ABC} \nabla^{C \dot B} W^2 \nonumber \\
&-& \frac{9}{16} i \alpha^2 \overline{W}^2 
\left(\overline{\nabla}^2 \overline{W}^2 \right) W_{ABC}
\nabla^B \nabla^C_{\ \ \dot A} W^2
\nonumber \\
&-& \frac{3}{4} i \alpha^2 \overline{W}^2
\left(\overline{\nabla}^2 \nabla^C_{\ \ \dot A} \overline{W}^2 \right) W_{ABC}
\nabla^B W^2 \nonumber \\
&-& \frac{9}{32} i \alpha^2 \left(\overline{\nabla}^2 \overline{W}^2 \right) 
W^2 W_{\dot A}^{\ \ \dot B \dot C} \nabla_{\dot B} \nabla_{A \dot C}  
\overline{W}^2 \nonumber \\
&+& \frac{9}{4} i \alpha^2 \left(\nabla^{\dot B} \overline{W}^2 \right)
W_{\underline{\dot B} \dot D \dot E} W_{ABC} \nabla^E 
\nabla_{E \underline{\dot A}} \nabla^{B \dot D} \nabla^{C \dot E} W^2
\nonumber \\
&-& \frac{9}{4} i \alpha^2 \left(\nabla^{\dot B} \overline{W}^2 \right)
W_{\dot A \dot B \dot C} W_{ABC} \nabla^B 
\nabla^{D \dot D} \nabla_{D \dot D} \nabla^{C \dot C} W^2
\nonumber \\
&+& \frac{9}{2} i \alpha^2 \left(\nabla^{\dot B} \overline{W}^2 \right)
\left(\nabla_{\underline{\dot A}}
W_{\underline{\dot B} \underline{\dot D} \underline{\dot E}} \right)
W^{\dot D \dot E \dot C} W_{ABC} \nabla^B \nabla^C_{\ \ \dot C} W^2
\nonumber \\
&+& \frac{9}{4} i \alpha^2 \left(\nabla^{\dot B} \overline{W}^2 \right)
W_{\dot A \dot B \dot C} W_{ABC} \left(\nabla^{\underline{B}}
W^{\underline{C} \underline{D} \underline{E}} \right) 
\nabla_D \nabla_E^{\ \ \dot C} W^2 \nonumber \\
&+& \frac{27}{4} \alpha^2 \left(\nabla^{B \dot B} \overline{W}^2 \right)
W_{\dot A \dot B \dot C} W_{ABC}
\nabla^{D \dot D} \nabla_{D \dot D} \nabla^{C \dot C} W^2
\nonumber \\
&-& \frac{9}{64} \alpha^2 \left(\overline{\nabla}^2 \overline{W}^2 \right)
W_{\dot A \dot B \dot C} W_{ABC} 
\nabla^2 \nabla^{B \dot B} \nabla^{C \dot C} W^2 
\nonumber \\
&+& \frac{27}{4} i \alpha^2 \left( \nabla_{\underline{\dot A}} 
\nabla^B_{\ \ \underline{\dot B}} \overline{W}^2 \right)
W_{\dot D}^{\ \ \dot B \dot C} W_{ABC}
\nabla^D \nabla_{D \dot C} \nabla^{B \dot D} W^2
\nonumber \\
&+& \frac{9}{2} i \alpha^2 \left( \nabla^{\dot D}
\nabla^B_{\ \ \dot D} \overline{W}^2 \right)
W_{\dot A \dot B \dot C} W_{ABC}
\nabla^D \nabla_D^{\ \ \dot C} \nabla^{B \dot B}  W^2
\nonumber \\
&-& \frac{27}{2} \alpha^2 \left(\nabla^{B \dot B} \overline{W}^2 \right)
\left(\nabla_{\underline{\dot A}}
W_{\underline{\dot B} \underline{\dot C} \underline{\dot D}} \right)
W^{\dot C \dot D \dot E} W_{ABC} \nabla^C_{\ \ \dot E} W^2
\nonumber \\
&+& \frac{9}{16} i \alpha^2 W^2 \left(\nabla^{\dot B} \overline{W}^2 \right)
\nabla_{\underline{\dot A}} \nabla^{D \dot D} \nabla_{D \dot D}
\nabla_{A \underline{\dot B}} \overline{W}^2
\nonumber \\
&+& \frac{9}{8} i \alpha^2 W^2 \left(\nabla^{\dot B} \overline{W}^2 \right)
\left(\nabla_{\underline{\dot A}}
W_{\underline{\dot B} \underline{\dot C} \underline{\dot D}} \right)
\nabla^{\dot D} \nabla_A^{\ \ \dot C} \overline{W}^2
\nonumber \\
&-& \frac{9}{64} i \alpha^2 W^2 
\left(\overline{\nabla}^2 \overline{W}^2 \right)
\nabla^{D \dot D} \nabla_{D \dot D} \nabla_{A \dot A} \overline{W}^2
\nonumber \\
&-& \frac{3}{2} i \alpha^2 \left(\overline{\nabla}^2 \overline{W}^2 \right)
\left( \nabla^C_{\ \ \dot A} \overline{W}^2 \right) W_{ABC} \nabla^B W^2
\nonumber \\
&-& \frac{27}{16} i \alpha^2 \left(\nabla^{\dot B} \overline{W}^2 \right)
\left(\nabla_{\dot B} \overline{W}^2 \right) W_{ABC}
\nabla^B \nabla_{\ \ \dot A}^C W^2
\nonumber \\
&-& \frac{9}{4} i \alpha^2 \left(\nabla^{\dot B} \overline{W}^2 \right)
\left( \nabla_{\underline{\dot A}} \nabla^B_{\ \ \underline{\dot B}} 
\overline{W}^2 \right) W_{ABC} \nabla^B W^2
\nonumber \\
&+& \frac{9}{2} i \alpha^2 \left(\nabla^{\dot B} \overline{W}^2 \right)
\left(\nabla^B_{\ \ \underline{\dot A}} W_{\underline{\dot B} 
\underline{\dot C} \underline{\dot D}} \right) W_{ABC}
\nabla^D \nabla_D^{\ \ \dot D} \nabla^{C \dot C} W^2
\nonumber \\
&-& \frac{9}{8} i \alpha^2 \left(\nabla^{\dot B} \overline{W}^2 \right)
\left(\nabla^{\dot D} \nabla_{D \dot D} \nabla_{E \dot A} \overline{W}^2  
\right) W_{ABC} \nabla^{\underline{B}}_{\ \ \dot B} 
W^{\underline{C} \underline{D} \underline{E}}
\nonumber \\ 
&-& \frac{15}{4} \alpha^2 \left(\nabla^{\dot B} \overline{W}^2 \right)
\left(\nabla^B_{\ \ \underline{\dot B}} \overline{W}^2 \right) W_{ABC}
\nabla^C_{\ \ \underline{\dot A}} W^2
\nonumber \\
&+& \frac{9}{16} \alpha^2 \left(\nabla^{\dot B} \overline{W}^2 \right)
\left(\overline{\nabla}^2 \nabla_{D \underline{\dot B}} 
\nabla_{E \underline{\dot A}} \overline{W}^2 \right)
W_{ABC} \nabla^{\underline{B}} W^{\underline{C} \underline{D} \underline{E}}
\nonumber \\
&-& \frac{27}{4} i\alpha^2 \left(\nabla^{B \dot B} \overline{W}^2 \right)
\left(\nabla_{\underline{\dot A}} W_{\underline{\dot B} 
\underline{\dot C} \underline{\dot D}} \right) W_{ABC}
\nabla^D \nabla_D^{\ \ \dot D} \nabla^{C \dot C} W^2
\nonumber \\
&-& \frac{9}{32} \alpha^2 \left(\nabla^{\dot B} \nabla_{D \dot B} 
\nabla_{E \dot A} \overline{W}^2 \right)
\left(\overline{\nabla}^2 \overline{W}^2 \right)
W_{ABC} \nabla^{\underline{B}} W^{\underline{C} \underline{D} \underline{E}}
\nonumber \\
&+&\frac{3}{8} i \alpha^2 \left(\nabla_{\dot A} \overline{W}^2 \right)
\left( \nabla^{\dot B} \nabla^B_{\ \ \dot B} \overline{W}^2 \right) 
W_{ABC} \nabla^C W^2 
\nonumber \\
&-& \frac{9}{8} i \alpha^2 \left(\nabla^{\dot B} \overline{W}^2 \right)
\left( \nabla_{\underline{\dot A}} \nabla^B_{\ \ \underline{\dot B}} 
\overline{W}^2 \right) W_{ABC} \nabla^C W^2
\nonumber \\ 
&+& \left. \frac{27}{8} \alpha^2 \left(\nabla^{B \dot B} \overline{W}^2 \right)
\left(\nabla_{\dot A} \overline{W}^2 \right) W_{ABC}
\nabla^C_{\ \ \dot B} W^2
+\mathrm{h.c.} \right] +{\cal O}\left( \alpha^3\right) \label{g2}
\end{eqnarray}

Substituting (\ref{g2}) and the exact expression (\ref{r}) in the expressions 
of appendix \ref{appendix1.2}, we get the solution of the superspace torsions 
and curvatures to second order in $\alpha$. The knowledge of these expressions
(though (\ref{g2}) being very complicated) may be relevant in order to check
consistency with higher dimensional results, such as the on-shell solution of 
Bianchi identities or the identification of auxiliary fields and constraints 
such as (\ref{diffg}).

\section{Discussion, conclusions and future perspectives}
\label{5}
\indent

In this paper, we have shown that, in order to supersymmetrize ${\cal R}^4$
on shell, one must introduce an infinite number of terms. We derived this 
result in four dimensions, but it is valid in higher dimensions. Any 
dimensional reduction of an equivalent higher-dimensional result must agree 
with it.
Its infinite number of terms shows that the complete supersymmetric 
${\cal R}^4$ theory is nonlocal.

By being able to get a complete solution for $R$ (\ref{r}), we see that we can put our theory only partially off-shell (by eliminating the auxiliary fields $M, N$ and leaving $A_m$) with just a finite number of terms. 

The result we obtained is not unexpected: since the late 80s, during the supersymmetrization of the Lorentz Chern-Simons term, people obtained nonlinear differential equations for auxiliary fields \cite{bss87, bpt87, rrz89}. The same was valid for the four-dimensional case \cite{cfgp85}.

In \cite{rrz89}, auxiliary fields $Y_{mnpq}$ and $S_{mnp}$ were introduced while solving the Bianchi identities. These fields had to be expanded to an infinite series in a perturbative parameter $c_2$ (defined below). What is remarkable is that a complete, all order solution to $Y_{mnpq}$ exists (but depending on $S_{mnp}$). It is then possible to eliminate $Y_{mnpq}$, leaving only $S_{mnp}$. This situation is similar to the one we saw in this paper: the auxiliary fields $Y_{mnpq}$ and $S_{mnp}$ are analogous to our $R$ and $A_m$, respectively.

There is, anyway, an important difference between working in ten or eleven dimensions and the four dimensional case we considered in this paper. In ten dimensional type I supergravity in superspace \cite{n81}, one must introduce a 2-form $B_{MN}$, with field strength $H_{MNP}$, in order to accomodate the $x$-space 2-form $B_{mn}$. The field strength $H$ satisfies the Bianchi identity $dH=0$ (in differential form notation) which, together with the torsion Bianchi identities, form a set of coupled equations to be solved simultaneously. If we consider type I supergravity coupled to super Yang-Mills \cite{nt86}, we have an additional one-form $A_M$ with field strength $F_{MN}$ and Bianchi identity $dF=0$, but we must modify the $H$ Bianchi identity by $dH=\mbox{tr} F^2$, because of the corresponding redefinition of the $x$-space field strength \cite{cm83}. In eleven-dimensional supergravity there is a similar situation \cite{cf80}: one needs to introduce a superspace 3-form $X_{MNP}$ in order to accomodate the $x$-space 3-form $X_{mnp}$, but its field strength $Y_{MNPQ}$ satisfies the Bianchi identity $dY=0$. In all these cases, even if we had an eventual off-shell formalism, we would still have a system of coupled nonlinear superspace Bianchi identities which would have to be solved by an iterative procedure. \footnote{If we instead consider anomaly-free ten-dimensional supergravity, there is an extra complication: the Green-Schwarz mechanism requires an extra redefinition of the $x$-space three-form field strength, as we mentioned, which implies an extra term in $dH=\mbox{tr} (c_1F^2 - c_2 R^2)$. This difficulty can be remedied with a theorem proved in \cite{bpt87}, which reduces this case to the previous one ($dH=\mbox{tr} F^2$) through a superfield redefinition.}

In four dimensional supergravity, because of its smaller field content, we only need to consider the torsion Bianchi identities. As it is well known, there is a complete off-shell solution to them. The non-linearities may only show up when we go on-shell, as we did with a higher-derivative action. This was the reason for the need of an infinite number of terms in a completely supersymmetric action. From the results of this paper, it should be very clear that the necessity for an infinite number of terms in the corresponding problem in ten or eleven dimensions comes also from the fact that we are supersymmetrizing a higher-derivative theory, and not only because of coupled nonlinear Bianchi identities.

Knowing this, there are two alternatives to proceed supersymmetrizing ${\cal R}^4$ actions. One is to work order by order in some perturbative parameter, changing the supersymmetry transformations in $x$-space. This is what has been done in \cite{dsw92} in ten dimensions and in \cite{pvw00} in eleven. The second alternative is to introduce auxiliary fields. They are easier to identify while solving the Bianchi identities in superspace, where one can relax the torsion constraints. This procedure has been followed by \cite{cgnn00, gn01, nr01}.

From the results we got in this paper, we can compare our solution to the superspace Bianchi identities with higher dimensional ones, after eliminating auxiliary fields in these solutions. A comparison of the solutions before and after the elimination of the auxiliary fields could clarify their role and possible identification.

We can also expand the action (\ref{action}) in components and eliminate the auxiliary fields $M, N$, getting a partially off-shell supersymmetric $x$-space ${\cal R}^4$ action, which should be easier to compare with the ${\cal R}^4$ superinvariants constructed in higher dimensions. This work is in progress \cite{m01}.

\paragraph{Acknowledgements}
\noindent
 
The author is grateful to Martin Ro\v cek for having suggested him the 
problem and for very helpful discussions, assistance and guidance. He is also grateful 
to Peter van Nieuwenhuizen for having spent many hours teaching him 
supergravity, and for very important help in crucial points of the calculation,
by having shown and explained him a draft of the still unpublished book 
\cite{west} with the derivation of (\ref{hbk}), (\ref{hadb}), (\ref{hkkhaa}) 
and (\ref{hkkhdada}). He also wants to thank Warren Siegel for very useful 
remarks.

This work has been supported by Funda\c c\~ao para a Ci\^encia e a
Tecnologia (Portugal) through grant PRAXIS XXI/BD/11170/97.


\appendix


\section{Survey of known results and conventions}
\setcounter{equation}{0}
\label{appendix1}


\subsection{General conventions}
\indent

We work in flat Minkowski space-time with the metric $\eta_{mn}={\mathrm diag} 
\left(-1, 1, 1, 1 \right)$. The Levi-Civita tensor is given by 
$\varepsilon^{0123}=-\varepsilon_{0123}=1$.

In flat space the vector indices are $m, n, \ldots$, the two-component spinor 
indices are $A, \dot A, \ldots$, and the superindices are $M, N, \ldots$ In 
curved space, the vector indices are $\mu, \nu, \ldots$ and the superindices 
are $\Lambda, \Pi, \ldots$ We do not use curved spinor indices in $x$-space.

We raise and lower two-component spinor indices with the tensors
\begin{equation}
\varepsilon_{AB}=\varepsilon^{AB}=-\varepsilon_{\dot A \dot B}=
-\varepsilon^{\dot A \dot B}, \, \varepsilon_{12}=1
\end{equation}
The contractions of spinor indices {\em always} follow the north-west rule 
(the same is valid for superindices in general). We then define
\begin{equation}
\theta^2=\theta^A \theta_A = \varepsilon^{AB} \theta_B \theta_A, \,
{\overline \theta}^2=\theta^{\dot A} \theta_{\dot A} = 
\varepsilon^{\dot A \dot B} \theta_{\dot B} \theta_{\dot A}
\end{equation}

The decomposition of a tensor with two spinor indices on its symmetric 
(underlined) and antisymmetric (trace) parts is always (for dotted/undotted, 
upper/lower indices) given by
\begin{equation}
T_{AB}=T_{\underline{A} \underline{B}}-\frac{1}{2} \varepsilon_{AB} 
T^C_{\ \ C}, \, \, \, T_{\dot A \dot B}=T_{\underline{\dot A} 
\underline{\dot B}}
-\frac{1}{2} \varepsilon_{\dot A \dot B} T^{\dot C}_{\ \ \dot C}
\end{equation}

Dotted and undotted spinor indices are related through complex conjugation
(that's what a bar means):
\begin{equation}
\overline{\theta_A}= \theta_{\dot A}, \, \, 
\overline{\theta^A}= -\theta^{\dot A}
\end{equation}
For fermionic derivatives, the rules are
\begin{equation}
\overline{\left( \frac{\partial}{\partial \theta_A} \right)}
=-\frac{\partial}{\partial \theta_{\dot A}}, \, \, 
\overline{ \left( \frac{\partial}{\partial \theta^A} \right)} =
\frac{\partial}{\partial \theta^{\dot A}}
\end{equation}

We define (as proposed in \cite{ggrs} - notice that our conventions differ 
from this book)
\begin{equation}
A_{A \dot A}= A_m \sigma^m_{\ \ A \dot A}, \, \, 
A_m = \frac{1}{2} A_{A \dot A} \sigma_m^{\ \ A \dot A}
\end{equation}
with
\begin{equation}
\sigma^m_{\ \ A \dot A}=\left(I, \vec{\sigma} \right)_{A \dot A}, \, \, 
\sigma_m^{\ \ A \dot A}=\left(-I, \vec{\sigma} \right)^{A \dot A}
\end{equation}
$\vec{\sigma}$ are the three Pauli matrices.

We write (in superspace, at least) all the vector indices contracted like 
this. Because of the Lorentz invariance of $\sigma^m_{\ \ A \dot A}$, we can 
use in our superspace calculations with these ``contracted vector indices'' 
the normal rules for the spinor indices.


\subsection{Superspace conventions}
\label{appendix1.2}
\indent

The superspace Lorentz covariant derivative is given by
\begin{equation}
\nabla_\Lambda =\partial_\Lambda +\frac{1}{2}\Omega_\Lambda^{\ \
mn}J_{mn}, \, \, \nabla_M=E_M^{\ \ \Lambda }\nabla_\Lambda
\end{equation} 

Our definitions of supertorsion and (Lorentz-valued) supercurvature come from
\begin{equation}
\left[ \nabla_M,\nabla_N \right\} =T_{MN}^{\ \ \ \ R}\nabla
_R+\frac{1}{2}R_{MN}^{\ \ \ \ rs}J_{rs}
\end{equation}
Explicitly, we have
\begin{eqnarray}
T_{MN}^{\ \ \ \ R}&=&E_M^{\ \ \Lambda }\partial_\Lambda E_N^{\ \ \Pi
}E_\Pi^{\ R}+\Omega_{MN}^{\ \ \ \ R}-\left( -\right)^{MN}\left(M
\leftrightarrow N\right) \label{tor} \\
R_{MN}^{\ \ \ \ rs}&=&E_M^{\ \ \Lambda }E_N^{\ \ \Pi }\left\{ \partial
_\Lambda \Omega _\Pi^{\ rs}+\Omega _\Lambda^{\ rk}\Omega_{\Pi k}^{\ \
\ s}-\left( -\right)^{\Lambda \Pi }\left( \Lambda \leftrightarrow \Pi
\right) \right\} 
\end{eqnarray}

Our choice of constraints is that all the torsions which are missing in the
following list are set equal to zero, with the exception of
\begin{equation}
T_{A\dot B}^{\ \ \ \ m}=-2i\sigma _{A\dot B}^m \label{tadbm}
\end{equation}

The solutions to the Bianchi identities \cite{gwz79} are, in our conventions, 
for the torsions: 
\begin{eqnarray}
T_{A\dot A\ \dot B C}&=&\frac{i}{12}\varepsilon_{AC}\varepsilon_{\dot A \dot
B}\overline{R} \label{tadadbc}
\\
T_{A \dot A\ BC}&=&\frac{i}{4}\left( 3\varepsilon _{AB}G_{C\dot
A}+\varepsilon_{AC}G_{B\dot A}-3\varepsilon_{BC}G_{A\dot A}\right) 
\label{tadabc}
\\
T_{A\dot A\ B\dot B\ C}&=&-\varepsilon_{\dot A\dot B}\left(
W_{ABC}-\frac{1}{2}\varepsilon_{AC} \nabla^{\dot C} G_{B\dot
C}-\frac{1}{2}\varepsilon_{BC} \nabla^{\dot C} G_{A\dot C}\right) \nonumber \\
&-&\varepsilon_{AB} \left( \ \nabla _{\underline{\dot A}} G_{C \underline{
\dot B}}\right) \label{tadabdbc}
\end{eqnarray}
Obviously, complex-conjugating (\ref{tadadbc}), (\ref{tadabc}), 
(\ref{tadabdbc}) we get the solutions for the complex-conjugated torsions. The
same is valid for the curvatures, whose solutions to the Bianchi identities 
are:

\begin{eqnarray}
R_{ABCD}&=&\frac{1}{6} \left( \varepsilon_{AC} \varepsilon_{BD}
+\varepsilon_{AD} \varepsilon_{BC}\right) R
\\
R_{A\dot B C D}&=&\varepsilon_{AC} G_{D\dot B}+\varepsilon_{AD}G_{C\dot B}
\\
R_{\dot A \dot B C D}&=&0
\\
R_{E\ A\dot A\ \dot B\dot C}&=&-2i\varepsilon_{EA} W_{\dot A\dot B\dot
C}-\frac{i}{96} \varepsilon_{EA} \varepsilon_{\dot A\dot B} \nabla_{\dot
C}R-\frac{i}{96} \varepsilon_{EA} \varepsilon_{\dot A\dot C} \nabla_{\dot B}
R \nonumber \\&+&\frac{i}{2} \varepsilon_{\dot A \dot C} \nabla_{\underline{E}}
G_{\underline{A} \dot B} +\frac{i}{2}\varepsilon_{\dot A \dot B}
\nabla_{\underline{E}} G_{\underline{A} \dot C}
\\
R_{E\ A\dot A\ BC}&=& -i\varepsilon_{EA} \nabla_{\underline{B}} 
G_{\underline{C} \dot A}
-\frac{i}{2} \varepsilon_{EB} \nabla_{\underline{C}} G_{\underline{A} \dot A} 
-\frac{i}{2} \varepsilon_{EC} \nabla_{\underline{A}} G_{\underline{B} \dot A} 
\nonumber \\
&-&\frac{i}{32} \left( \varepsilon_{EB} \varepsilon_{AC}+ \varepsilon_{EC}
\varepsilon_{AB} \right) \nabla_{\dot A}R
\\
R_{A\dot A\ B\dot B\ CD}
&=& -\varepsilon_{\dot A \dot B} \nabla_{\underline{A}} W_{\underline{BCD}}
\nonumber \\
&-&\varepsilon_{\dot A \dot B} \left( \varepsilon_{AC}
\varepsilon_{BD} +\varepsilon_{AD} \varepsilon_{BC} \right) \cdot \nonumber \\ 
&\left. \right.& \left( \frac{1}{192} \left( \nabla^2 \overline{R} 
+\overline{\nabla }^2 R\right)
+\frac{1}{144} R \overline{R} 
+\frac{1}{4} G^{E\dot E} G_{E\dot E} \right)   \nonumber \\
&+& \varepsilon_{AB} \left[ \nabla_{\underline{C}} \nabla_{\underline{\dot A}}
G_{\underline{D \dot B}} -i \nabla_{\underline{C \dot A}} G_{\underline{D
\dot B}} +G_{\underline{C \dot A}} G_{\underline{D\dot B}}\right]
\end{eqnarray}

The superfields $G_{A \dot A}$ and $W_{ABC}$ have the following complex 
conjugation properties:

\begin{equation}
\overline{G_{A\dot B}}=G_{B\dot A}
\end{equation} 
\begin{equation}
\overline{W_{ABC}}=W_{\dot A\dot B\dot C}
\end{equation}

$R$ and $W_{\dot A\dot B\dot C}$ are antichiral:

\begin{eqnarray}
&\left. \right.&
\nabla_A R=0 \\
&\left. \right.&
\nabla_A W_{\dot A\dot B\dot C}=0
\end{eqnarray}

The following relations between $R$, $G_{A \dot A}$ and $W_{ABC}$ are a
consequence of the Bianchi identities:
\begin{equation}
\nabla^A G_{A\dot B} = \frac{1}{24} \nabla_{\dot B} R  \label{diffg}
\end{equation}

\begin{equation}
\nabla^A W_{ABC} = i \left( \nabla_{B \dot A} G_C^{\ \ \dot A} +\nabla_{C
\dot A} 
G_B^{\ \ \dot A} \right) 
\label{diffw}
\end{equation}

From (\ref{diffg}) and its complex conjugate and the solution of the Bianchi 
identities, we may also derive the following useful relation between 
superfields:

\begin{equation}
\nabla^2 \overline{R} - \overline{\nabla}^2 R = 96 i \nabla^n G_n 
\label{wb203}
\end{equation}
These relations (\ref{diffg}), (\ref{diffw}), (\ref{wb203}) are off-shell 
identities (not field equations).


\section{The constrained variation of $W_{ABC}$}
\label{b}
\setcounter{equation}{0}
\indent

In this appendix, we present the details of the computation of $\delta W_{ABC}$.
From the solution of the Bianchi identities, we have

\begin{equation}
W_{ABC} = \frac{1}{2} T_{\underline{A}\ \ \ \underline{B} \dot A \
\underline{C}}^{\ \ \dot A}, \, \, 
W_{\dot A \dot B \dot C} = -\frac{1}{2} T_{\ \ \underline{\dot A} A
\underline{\dot B} \underline{\dot C}}^A
\end{equation}
Therefore, from (\ref{deltatmnr}) and the torsions (\ref{tadbm}), 
(\ref{tadadbc}), (\ref{tadabc}) and (\ref{tadabdbc}), we have for the 
constrained variation

\begin{eqnarray}
\delta W_{ABC}
&=&-H_{\underline{A}}^{\ \ \dot A\ D} T_{D\ \underline{B}\dot
A\ \underline{C}}
-\frac{1}{2} H_{\underline{A}}^{\ \ \dot A\ D\dot D}T_{D\dot D\ 
\underline{B}\dot A\ \underline{C}} \nonumber \\
&+&\frac{1}{2} T_{\underline{A}\ \ \ \underline{B}\dot A\ }^{\ \ \dot A
\ \ \ \ D} 
H_{D \underline{C}} 
+\frac{1}{2} T_{\underline{A}\ \ \underline{B}\dot
A\ }^{\ \ \dot A\ \ \ \dot D} H_{\dot D\underline{C}}
-\nabla_{\underline{A}}^{\ \ \dot A} H_{\underline{B}\dot A\ \underline{C}}
\label{deltawabc}
\end{eqnarray}
This expression needs to be simplified. For that, we use again our knowledge of
the torsions from the solution of the Bianchi identities
and the equations for $H_M^{\ \ N}$ we have just 
derived. We get for each term in (\ref{deltawabc})

\begin{eqnarray}
- H_{\underline{A}}^{\ \ \dot A\ D}T_{D\ \underline{B}\dot A\ \underline{C}}
&=& \frac{3}{32}i \overline{R} \left( \nabla_{\underline{A}}
\chi_{\underline{B}}^{\ \ \dot A} \right) G_{\underline{C}\dot A} \nonumber \\
&+&\frac{3}{16}i \left( \nabla^{\dot A} \left( \nabla^{\dot E}
\nabla_{\underline{A}}- \nabla_{\underline{A}} \nabla^{\dot E} \right)
\chi_{\underline{B}\dot E}\right) G_{\underline{C} \dot A} \nonumber \\
&+&\frac{3}{8}\left( \nabla_{\underline{A}}^{\ \ \dot E} \nabla^{\dot A}
\chi_{\underline{B} \dot E} \right) G_{\underline{C}\dot A} \nonumber \\
&+& \frac{i}{16}\chi_{\underline{B}}^{\ \ \dot A}
\left( \nabla_{\underline{A}} \overline{R}\right) G_{\underline{C} \dot A} 
-\frac{3}{16} i G_{\underline{A} \dot E} G_{\underline{C} \dot A}
\nabla^{\dot E} \chi_{\underline{B}}^{\ \ \dot A}
\end{eqnarray}

\begin{eqnarray}
H_{\underline{A}}^{\ \ \dot A\ D\dot D}T_{D\dot D\
\underline{B}\dot A\ \underline{C}} 
&=&\left[ \frac{2}{3}\left( \nabla^2+\frac{1}{3}R\right) U 
+\frac{2}{3}\left( \overline{\nabla}^2 
+\frac{1}{3}\overline{R}\right) \overline{U} \right. \nonumber \\
&+& \left. \frac{4}{3}i \chi^{E\dot E} G_{E\dot E}
+\frac{i}{3}\left( \nabla^E \nabla^{\dot E}-\nabla^{\dot E} \nabla^E \right) 
\chi _{E\dot E}\right] W_{ABC} \nonumber \\
&+& \frac{i}{2}\left( \nabla_{\dot A} \nabla_{\underline{A}}-
\nabla_{\underline{A}}\nabla _{\dot A}\right) \chi^{D\dot A}W_{D\underline{BC}}
\nonumber \\
&-&\frac{i}{96}\left[ \left( \nabla^{\dot A}\nabla_{\underline{A}} 
-\nabla _{\underline{A}}\nabla^{\dot A}\right) \chi_{\underline{C}\dot A}
\right] \nabla_{\underline{B}}\overline{R} \nonumber \\
&+&\frac{i}{2} \left[ \left( \nabla^{\dot
A}\nabla_{\underline{A}}-\nabla_{\underline{A}} \nabla^{\dot A}\right)
\chi_{\underline{B}\dot D}\right]
\nabla^{\underline{\dot D}}G_{\underline{C}\underline{\dot A}}
\end{eqnarray}

\begin{eqnarray}
T_{\underline{A}\ \ \underline{B}\dot A}^{\ \ \dot A\ \ \ \ D}
H_{D\underline{C}}
&=&\left[ \frac{2}{3} \left( \nabla^2+\frac{1}{3} R\right) U
-\frac{1}{3}\left(
\overline{\nabla }^2+\frac{1}{3}\overline{R}\right) \overline{U} 
+\frac{1}{2}\nabla _{E\dot E}\chi^{E\dot E} \right.
\nonumber \\
&+&  \frac{i}{3}\chi^{E\dot E}G_{E\dot E} +\left. \frac{i}{2}
\left( \nabla^E\nabla^{\dot E}-\nabla^{\dot E}\nabla^E\right) \chi_{E\dot
E}\right] W_{ABC}
\end{eqnarray}

\begin{equation}
T_{\underline{A}\ \ \underline{B}\dot A}^{\ \ \dot A\ \ \ \dot D} 
H_{\dot D\underline{C}}
=-\frac{i}{12} \left( \nabla_{\underline{A}} G_{\underline{B}}^{\ \ \dot D}
\right) \overline{R}\chi_{\underline{C}\dot D}
\end{equation}

\begin{eqnarray}
\nabla_{\underline{A}}^{\ \ \dot A} H_{\underline{B}\dot A\ \underline{C}}
&=& -\frac{1}{16}\nabla_{\underline{A}}^{\ \ \dot A}\left(\overline{R} 
\nabla_{\underline{B}}\chi_{\underline{C}\dot A}\right) 
-\frac{1}{24}\nabla_{\underline{A}}^{\ \ \dot A}\left(
\chi_{\underline{B}\dot A}\nabla_{\underline{C}}\overline{R}\right) 
\nonumber \\
&+& \frac{1}{8}\nabla_{\underline{A}}^{\ \ \dot A}\nabla_{\dot
A}\left( \nabla_{\underline{B}}\nabla^{\dot E}-\nabla^{\dot
E}\nabla_{\underline{B}}\right) \chi_{\underline{C}\dot E} \nonumber \\
&+&\frac{i}{4}\nabla_{\underline{A}}^{\ \ \dot
A}\nabla_{\underline{B}}^{\ \ \dot E}\nabla_{\dot A}
\chi_{\underline{C}\dot E} 
+\frac{3}{16}\nabla_{\underline{A}}^{\ \ \dot A}\left(
G_{\underline{B}}^{\ \ \dot E}\nabla_{\dot E}\chi_{\underline{C}\dot A}\right) 
\nonumber \\
&+&\frac{1}{16}\nabla_{\underline{A}}^{\ \ \dot A}\left(G_{\underline{B} \dot A} \nabla_{\dot E}\chi_{\underline{C}}^{\ \ \dot E}\right) 
-\frac{5}{16}\nabla_{\underline{A}}^{\ \ \dot A} 
\left(G_{\underline{B}}^{\ \ \dot E} 
\nabla_{\dot A}\chi_{\underline{C} \dot E}\right) 
\end{eqnarray}

We may then write (\ref{deltawabc}) as
\begin{eqnarray}
\delta W_{ABC}
&=&-\frac{1}{2}\left[ \left( \overline{\nabla }^2 
+\frac{1}{3}\overline{R}\right) \overline{U}\right] W_{ABC}
-\frac{i}{2}\chi^{E\dot E}G_{E\dot E}W_{ABC} \nonumber \\
&-& \frac{i}{8}\left[ \left( \nabla^E \nabla^{\dot E}-\nabla^{\dot E}
\nabla^E\right) \chi_{E\dot E}\right] W_{ABC} \nonumber \\
&-&\frac{i}{4}\left( \nabla_{\dot A}\nabla_{\underline{A}}-
\nabla_{\underline{A}}\nabla_{\dot A}\right) 
\chi^{D\dot A}W_{D\underline{BC}} \nonumber \\
&+&\frac{1}{4}\left( \nabla_{E\dot E}\chi^{E\dot E}\right)
W_{ABC} 
-\frac{i}{4}\left[ \left( \nabla^{\dot
A}\nabla_{\underline{A}} -\nabla_{\underline{A}}\nabla^{\dot A}\right)
\chi_{\underline{B}\dot D}\right] \nabla^{\dot D} G_{\underline{C} \dot A} 
\nonumber \\
&+& \frac{1}{16}\nabla_{\underline{A}}^{\ \ \dot A}\left(
\overline{R}\nabla_{\underline{B}}\chi_{\underline{C}\dot A}\right) 
-\frac{1}{8}\nabla_{\underline{A}}^{\ \ \dot A}\nabla_{\dot
A}\left( \nabla_{\underline{B}}\nabla^{\dot E}-\nabla^{\dot
E}\nabla_{\underline{B}}\right) \chi_{\underline{C}\dot E} \nonumber \\
&-&\frac{i}{4}\nabla_{\underline{A}}^{\ \ \dot
A}\nabla_{\underline{B}}^{\ \ \dot E}\nabla_{\dot
A}\chi_{\underline{C} \dot E} 
+\frac{1}{24}\nabla_{\underline{A}}^{\ \
\dot A}\left( \chi_{\underline{B}\dot
A}\nabla_{\underline{C}}\overline{R}\right) \nonumber \\
&-& \frac{3}{16}\nabla_{\underline{A}}^{\ \ \dot
A}\left(G_{\underline{B}}^{\ \ \dot E}\nabla_{\dot
E}\chi_{\underline{C}\dot A}\right) 
- \frac{1}{16}\nabla_{\underline{A}}^{\ \ \dot A}
\left( G_{\underline{B}\dot A}\nabla_{\dot
E}\chi_{\underline{C}}^{\ \ \dot E}\right) \nonumber \\
&+& \frac{5}{16}\nabla_{\underline{A}}^{\ \
\dot A}\left( G_{\underline{B}}^{\ \ \dot E}\nabla_{\dot
A}\chi_{\underline{C}\dot E}\right) 
-\frac{i}{24}\left(
\nabla_{\underline{A}}G_{\underline{B}}^{\ \ \dot D}\right)
\overline{R}\chi_{\underline{C}\dot D} \nonumber \\
&+& \frac{3}{32}i\overline{R}\left(
\nabla_{\underline{A}}\chi_{\underline{B}}^{\ \ \dot A}\right)
G_{\underline{C}\dot A} 
+\frac{i}{16}\chi_{\underline{B}}^{\ \ \dot A}\left(
\nabla_{\underline{A}}\overline{R}\right) G_{\underline{C}\dot A} \nonumber \\
&+& \frac{3}{16}i\left( \nabla^{\dot A}\left(
\nabla^{\dot E}\nabla_{\underline{A}}-\nabla_{\underline{A}}\nabla^{\dot
E}\right) \chi_{\underline{B}\dot E}\right) G_{\underline{C}\dot A} 
\nonumber \\
&+& \frac{3}{8}\left( \nabla_{\underline{A}}^{\ \ \dot E}\nabla^{\dot
A}\chi_{\underline{B}\dot E}\right) G_{\underline{C}\dot A} 
-\frac{3}{16}iG_{\underline{A}\dot E} G_{\underline{C}\dot A}\nabla^{\dot
E} \chi_{\underline{B}}^{\ \ \dot A}
\end{eqnarray}
and hence

\begin{eqnarray}
2\overline{W}^2W^{ABC}\delta W_{ABC} &=& 
-i\chi^{E\dot E}G_{E\dot E} W^2\overline{W}^2 
+\frac{1}{2}\left( \nabla_{E\dot E}\chi^{E\dot E}\right) W^2\overline{W}^2 \nonumber \\
&-& \frac{i}{2}\left[ \left( \nabla^{\dot
A}\nabla_A-\nabla_A\nabla^{\dot A}\right) \chi_{B\dot D}\right] \left(
\nabla^{\dot D}G_{C \dot A}\right) W^{ABC}\overline{W}^2 \nonumber \\
&-& \frac{1}{8}\nabla_{A}^{\ \ \dot A}\left(
\overline{R}\nabla_B\chi_{C\dot A}\right) W^{ABC} \overline{W}^2 \nonumber \\
&+& \frac{1}{4}\left[ \nabla_{A}^{\ \ \dot
A}\nabla_{\dot A}\left( \nabla_B\nabla^{\dot E}-\nabla^{\dot E}
\nabla_B\right) \chi_{C\dot E}\right] W^{ABC}\overline{W}^2 \nonumber \\
&+& \frac{i}{2}\left[ \nabla_{A}^{\ \ \dot A}\nabla_{B}^{\
\ \dot E}\nabla_{\dot A}\chi_{C\ \dot E}\right]
W^{ABC}\overline{W}^2 \nonumber \\
&-& \frac{1}{12}\left[ \nabla_{A}^{\ \ \dot
A}\left( \chi_{B\dot A} \nabla_C\overline{R}\right) \right] 
W^{ABC}\overline{W}^2 \nonumber \\
&+& \frac{3}{8} \left[ \nabla_{A}^{\ \ \dot A}
\left( G_{B}^{\ \ \dot E}\nabla_{\dot
E}\chi_{C\dot A}\right) \right] W^{ABC}\overline{W}^2 \nonumber \\
&+& \frac{1}{8}\nabla_{A}^{\ \ \dot A}\left( G_{B\dot A}\nabla_{\dot
E}\chi_{C}^{\ \ \dot E}\right) W^{ABC}\overline{W}^2 \nonumber \\
&-& \frac{5}{8}\nabla_{A}^{\ \ \dot A}\left(
G_{B}^{\ \ \dot E}\nabla_{\dot A}\chi_{C\ \dot E}\right) W^{ABC}\overline{W}^2
\nonumber \\
&+& \frac{i}{12}\left( \nabla_A G_B^{\ \ \dot D}\right)
\overline{R}\chi_{C\dot D}W^{ABC}\overline{W}^2 \nonumber \\
&-& \frac{3}{32}i\overline{R}\left( \nabla_A \chi_B^{\ \ \dot A}\right)
G_{C\dot A} W^{ABC}\overline{W}^2 \nonumber \\
&-& \frac{3}{8} i\left( \nabla^{\dot A}\left( \nabla^{\dot
E}\nabla_A-\nabla_A \nabla^{\dot E} \right) \chi_{B\dot E}\right) G_{C\dot
A} W^{ABC}\overline{W}^2 \nonumber \\
&-& \frac{3}{4}\left( \nabla_A^{\ \ \dot E}\nabla^{\dot A}\chi_{B\dot E}\right)
G_{C\dot A}W^{ABC}\overline{W}^2 \nonumber \\
&-& \frac{i}{8}\chi_B^{\ \ \dot A}\left( \nabla_A
\overline{R}\right) G_{C\dot A}W^{ABC} \overline{W}^2 \nonumber \\
&+& \frac{3}{8}i G_{A\dot
E}G_{C\dot A}\nabla^{\dot E}\chi_B^{\ \ \dot A}W^{ABC}\overline{W}^2 
\nonumber \\
&-& \left[ \left( \overline{\nabla }^2 +\frac{1}{3}\overline{R}\right)
\overline{U}\right] W^2\overline{W}^2 \label{2wdw}
\end{eqnarray}


\section{Exact computations for the simplification of $G_{A \dot A}$}
\label{c}
\setcounter{equation}{0}
\indent

Here we present the exact expansion of all the terms of (\ref{gcomplicated}).

\begin{eqnarray}
&\left. \right.& \left( \nabla_{\dot A}\nabla^B - \nabla^B \nabla_{\dot A}
\right) \nabla^{\dot D} \left( G_{\ \ \dot D}^C W_{ABC} \overline{W}^2\right)
\nonumber \\
&=&-\frac{i}{6} \overline{W}^2 \left( \nabla^B W_{ABC} \right) \nabla_{\ \
\dot A}^C \overline{R} 
+ \frac{i}{12} \overline{W}^2 \nabla^B W_{ABC}
\nabla^B \nabla_{\ \ \dot A}^C \overline{R} \nonumber \\
&-& \frac{5}{24} \overline{W}^2 W_{ABC} G_{\ \ \dot
A}^C \nabla^B \overline{R} 
+\frac{1}{24} \left(\nabla_{\dot A} \overline{W}^2 \right)
\left(\nabla^B W_{ABC} \right) \nabla^C \overline{R} \nonumber \\
&-& \frac{i}{12} \overline{W}^2 \left(\nabla_{\ \ \dot A}^C W_{ABC}\right) 
\nabla^B \overline{R} 
-2i \left(\nabla^{\dot D} \overline{W}^2 \right) W_{ABC}
\left(\nabla_{\ \ \underline{\dot A}}^B G_{\ \ \underline{\dot D}}^C
\right) \nonumber \\
&-& 2\left( \nabla^{\dot D} \overline{W}^2 \right) W_{ABC}
\left(\nabla^B \nabla_{\underline{\dot A}}G_{\ \ \underline{\dot D}}^C
\right) 
- 9 G_{\ \ \underline{\dot A}}^B G_{\ \ \underline{\dot D}}^C 
\left( \nabla^{\dot D} \overline{W}^2\right) W_{ABC} \nonumber \\
&+& 4i\left(\nabla^{B\dot D} \overline{W}^2\right) W_{ABC}
\nabla_{\underline{\dot A}} G_{\ \ \underline{\dot D}}^C 
+2\left( \nabla^{\dot D}\overline{W}^2\right) \left( \nabla^B W_{ABC}\right)
\nabla_{\underline{\dot A}} G_{\ \ \underline{\dot D}}^C \nonumber \\
&-& \left(\overline{\nabla}^2 \overline{W}^2 \right) W_{ABC} \nabla^B G_{\ \
\dot A}^C 
+ \left( \overline{\nabla}^2 \overline{W}^2 \right) \left(
\nabla^B W_{ABC}\right) G_{\ \ \dot A}^C \nonumber \\
&+& 3iG_{\ \ \dot A}^C \left(\nabla^{\dot D} \nabla_{\ \ \dot D}^C 
\overline{W}^2\right) W_{ABC}
+2i G^{C\dot D} \left(\nabla_{\underline{\dot D}} \overline{W}^2 \right)
\nabla_{\ \ \underline{\dot A}}^C W_{ABC}  \nonumber \\
&+& 2i G^{C\dot D}
\left(\nabla_{\underline{\dot D}} \nabla_{\ \ \underline{\dot A}}^C
\overline{W}^2 \right) W_{ABC}
+ \frac{1}{2}\left(\nabla_{\dot A}
\overline{W}^2 \right) W_A^{\ \ BC} \nabla^D W_{DBC}  \nonumber \\
&-& i G_{\ \ \dot A}^C
\left(\nabla^{\dot D} \overline{W}^2 \right) \nabla_{\ \ \dot D}^C W_{ABC}
\end{eqnarray}

\begin{eqnarray}
&\left. \right.&
\left(\nabla_{\dot D}\nabla^B- \nabla^B \nabla_{\dot D}\right) \left(
\left( \nabla_{\dot A} G^{C \dot D}\right) W_{ABC}\overline{W}^2\right)
\nonumber \\
&=&-\frac{i}{6} \overline{W}^2 \left( \nabla^B \nabla_{\ \ \dot A}^C
\overline{R} \right) W_{ABC} 
- \frac{i}{6} \overline{W}^2 \left( \nabla_{\
\ \dot A}^C \overline{R} \right) \nabla^B W_{ABC} \nonumber \\
&-& \frac{15}{16}\overline{W}^2 \left(\nabla^B \overline{R} \right) G_{\ \
\dot A}^C W_{ABC} 
+\frac{1}{4} \left(\nabla_{\dot A} \overline{W}^2 \right)
\left(\nabla^B W_{ABC} \right) \nabla^C \overline{R} \nonumber \\
&-& \frac{i}{24} \overline{W}^2 \left(\nabla^B \overline{R} \right) 
\nabla_{\ \ \dot A}^C W_{ABC}
+ \frac{i}{24} \left(\nabla^B \overline{R} \right) \left(\nabla_{\ \
\dot A}^C \overline{W}^2 \right) W_{ABC} \nonumber \\
&-& \overline{R} \overline{W}^2 G_{\ \ \dot A}^C \nabla^B W_{ABC} 
-\frac{5}{6} \overline{R} \overline{W}^2 \left(
\nabla^B G_{\ \ \dot A}^C \right) W_{ABC} 
- \frac{1}{2} W^2 \overline{W}^2 G_{A\dot A} \nonumber \\
&-& 3G^{B\dot D} \left( \nabla_{\underline{\dot A}} G_{\ \
\underline{\dot D}}^C \right) W_{ABC} \overline{W}^2 
+ 2i \overline{W}^2 \left( \nabla_{\underline{\dot A}} 
G_{\ \ \underline{\dot D}}^C \right) \nabla^{B\dot D}W_{ABC} \nonumber \\
&-& 2i \left ( \nabla_{\underline{\dot A}}G_{\ \ \underline{\dot D}}^C \right) 
\left( \nabla^{B\dot D} \overline{W}^2 \right) W_{ABC} 
- 2\left( \nabla_{\underline{\dot A}} G_{\ \ \underline{\dot
D}}^C \right) \left( \nabla^{\dot D} \overline{W}^2 \right) \nabla^B W_{ABC} 
\nonumber \\
&-&2 \left( \nabla^B \nabla_{\underline{\dot A}} G_{\ \ \underline{\dot D}}^C 
\right) \left( \nabla^{\dot D} \overline{W}^2\right) W_{ABC} 
+ 2i\overline{W}^2 W^{BCD} \nabla_{D\dot A}W_{ABC}
\end{eqnarray}

\begin{eqnarray}
&\left. \right.&
\nabla^{\dot D} \left(G_{\ \ \dot A}^B G_{\ \ \dot D}^C W_{ABC}
\overline{W}^2 \right) \nonumber \\
&=& -\frac{1}{16} \overline{W}^2 \left(\nabla^B
\overline{R} \right) G_{\ \ \dot A}^C W_{ABC} 
- G^{B\dot D} \left(\nabla_{\underline{\dot A}} G_{\ \ \underline{\dot D}}^C \right) W_{ABC} \overline{W}^2 \nonumber \\
&+& G_{\ \ \underline{\dot A}}^B G_{\ \ \underline{\dot
D}}^C \left( \nabla^{\dot D} \overline{W}^2\right) W_{ABC}
\end{eqnarray}

\begin{eqnarray}
&\left. \right.&
\nabla^{\dot D} \nabla_{\ \ \dot A}^B \left( G_{\ \ \dot D}^C W_{ABC}
\overline{W}^2 \right) \nonumber \\
&=&
-\frac{1}{24} \overline{W}^2 \left( \nabla^B
\nabla_{\ \ \dot A}^C \overline{R}\right) W_{ABC} 
+ \frac{11}{96} i\overline{W}^2 \left(\nabla^B \overline{R}\right)
G_{\ \ \dot A}^C W_{ABC} \nonumber \\
&-& \frac{1}{24} \overline{W}^2 \left(\nabla^B
\overline{R} \right) \nabla_{\ \ \dot A}^C W_{ABC} 
-\frac{1}{24} \left(\nabla^B \overline{R} \right) \left( \nabla_{\ \ \dot
A}^C \overline{W}^2 \right) W_{ABC} \nonumber \\
&-&\frac{i}{12}\overline{R} \left(
\nabla^B G_{\ \ \dot A}^C \right) \overline{W}^2 W_{ABC} 
-\frac{i}{12} \overline{R} G_{\ \ \dot
A}^C \overline{W}^2 \nabla^B W_{ABC} \nonumber \\
&+& \frac{3}{2} i G^{B\dot D} \left( \nabla_{\underline{\dot A}} 
G_{\ \ \underline{\dot D}}^C \right) W_{ABC} \overline{W}^2 
-G^{B\dot D} \left( \nabla_{\underline{\dot A}} \nabla_{\ \
\underline{\dot D}}^C \overline{W}^2 \right) W_{ABC} \nonumber \\
&-& G^{B\dot D} \left( \nabla_{\underline{\dot
A}} \overline{W}^2 \right) \nabla_{\ \ \underline{\dot D}}^C W_{ABC} 
+\frac{1}{2} G_{\ \ \dot A}^B \left( \nabla^{\dot D} \overline{W}^2 \right) 
\nabla_{\ \ \dot D}^C W_{ABC} \nonumber \\
&+& \frac{1}{2} G_{\ \ \dot A}^B \left( \nabla^{\dot D} \nabla_{\
\ \dot D}^C \overline{W}^2 \right) W_{ABC} 
-i G_{A\dot A} W^2 \overline{W}^2 \nonumber \\
&+&\left( \nabla_{\ \ \underline{\dot A}}^B G_{\ \ \underline{\dot D}}^C 
\right) \left( \nabla^{\dot D} \overline{W}^2 \right) W_{ABC} 
+\frac{i}{4} \left( \nabla_{\dot A} \overline{W}^2 \right) W_A^{\ \
BC} \nabla^D W_{DBC}
\end{eqnarray}

\begin{eqnarray}
&\left. \right.&
\nabla^B \left( \overline{R} \nabla_{\ \ \dot A}^C \left( W_{ABC}
\overline{W}^2 \right) \right) \nonumber \\
&=&\overline{W}^2 \left( \nabla^B \overline{R}
\right) \nabla_{\ \ \dot A}^C W_{ABC} 
+\left( \nabla^B \overline{R} \right)
\left( \nabla_{\ \ \dot A}^C \overline{W}^2 \right) W_{ABC} \nonumber \\
&+&\overline{R} \overline{W}^2 \nabla^B \nabla_{\ \ \dot A}^C W_{ABC}
+\overline{R} \left( \nabla_{\ \ \dot A}^C \overline{W}^2 \right) \nabla^B
W_{ABC}
\end{eqnarray}

\begin{eqnarray}
&\left. \right.&
\nabla^{\dot D} \left( G_{\ \ \dot D}^C \nabla_{\ \ \dot A}^B \left(W_{ABC}
\overline{W}^2 \right) \right) \nonumber \\
&=& -\frac{1}{24}\left( \nabla^B \overline{R}
\right) \left( \nabla_{\ \ \dot A}^C \overline{W}^2\right) W_{ABC} 
-\frac{1}{24} \overline{W}^2 \left( \nabla^B \overline{R} \right) \nabla_{\
\ \dot A}^C W_{ABC} \nonumber \\
&-& G^{B\dot D} \left( \nabla_{\underline{\dot A}}
\nabla_{\ \ \underline{\dot D}}^C \overline{W}^2 \right) W_{ABC} 
+\frac{1}{2} G_{\ \ \dot A}^B \left( \nabla ^{\dot D} \nabla_{\ \ \dot D}^C
\overline{W}^2 \right) W_{ABC} \nonumber \\
&-& G^{B\dot D} \left( \nabla_{\underline{\dot A}} \overline{W}^2 \right) 
\nabla_{\ \ \underline{\dot D}}^C W_{ABC} 
+\frac{1}{2} G_{\ \ \dot A}^B \left( \nabla^{\dot D} \overline{W}^2\right)
\nabla_{\ \ \dot D}^C W_{ABC} \nonumber \\
&-& \frac{i}{12} \overline{R} G_{\ \ \dot A}^C
\overline{W}^2 \nabla^B W_{ABC} 
- \frac{5}{96} i G_{\ \ \dot A}^C
\overline{W}^2 \left( \nabla ^B \overline{R} \right) W_{ABC} \nonumber \\
&-&\frac{5}{2} i G^{B\dot D} \left( \nabla_{\underline{\dot A}} 
G_{\ \ \underline{\dot D}}^C \right) W_{ABC} \overline{W}^2
\end{eqnarray}

\begin{eqnarray}
&\left. \right.&
\nabla_{\dot A} \left( G_{\ \ \dot D}^C \nabla^{B\dot D} \left(W_{ABC}
\overline{W}^2 \right) \right) \nonumber \\
&=&\left( \nabla_{\underline{\dot A}} G_{\ \ \underline{\dot D}}^C \right) 
W_{ABC} \nabla^{B \dot D} \overline{W}^2 
+ \overline{W}^2 \left( \nabla_{\underline{\dot A}} G_{\ \
\underline{\dot D}}^C \right) \nabla^{B \dot D} W_{ABC}
\end{eqnarray}

\begin{eqnarray}
&\left. \right.&
\nabla^{\dot D} \left( G_{\ \ \dot A}^B \nabla_{\ \ \dot D}^C \left(W_{ABC}
\overline{W}^2 \right) \right) \nonumber \\
&=&-\frac{1}{48}\left( \nabla^B \overline{R}
\right) \left( \nabla_{\ \ \dot A}^C \overline{W}^2 \right) W_{ABC} 
- \frac{1}{48} \left( \nabla^B \overline{R} \right) \overline{W}^2 \nabla_{\
\ \dot A}^C W_{ABC} \nonumber \\
&-& \frac{5}{48} i \left( \nabla^B \overline{R} \right)
G_{\ \ \dot A}^C \overline{W}^2 W_{ABC} 
- \frac{i}{6} \overline{R} G_{\ \ \dot A}^C \overline{W}^2 \nabla^BW_{ABC} 
\nonumber \\
&-& \left( \nabla_{\underline{\dot A}} G_{\ \
\underline{\dot D}}^C \right) W_{ABC} \nabla^{B\dot D} \overline{W}^2 
- \overline{W}^2 \left( \nabla_{\underline{\dot A}}G_{\ \ \underline{\dot
D}}^C \right) \nabla^{B\dot D} W_{ABC} \nonumber \\
&+& G_{\ \ \dot A}^B \left(
\nabla^{\dot D} \overline{W}^2 \right) \nabla_{\ \ \dot D}^C W_{ABC} 
+ G_{\ \ \dot A}^B \left( \nabla^{\dot D} \nabla_{\ \ \dot D}^C \overline{W}^2
\right) W_{ABC}
\end{eqnarray}

\begin{eqnarray}
&\left. \right.&
\nabla^{\dot D} \nabla _{\ \ \dot A}^B \nabla_{\ \ \dot D}^C \left(W_{ABC}
\overline{W}^2 \right) \nonumber \\ 
&=& -\frac{5}{48} i \overline{W}^2 \left(\nabla^B \nabla_{\ \ \dot A}^C 
\overline{R} \right) W_{ABC}  
- \frac{i}{6} \overline{W}^2 \left( \nabla_{\ \ \dot A}^C \overline{R} \right) 
\nabla^B W_{ABC} \nonumber \\
&-& \frac{5}{32} \overline{W}^2 G_{\ \ \dot A}^C \left( \nabla^B \overline{R} 
\right) W_{ABC} 
-\frac{5}{32} i \left( \nabla^B \overline{R} \right) \left( 
\nabla_{\ \ \dot A}^C \overline{W}^2 \right) W_{ABC} \nonumber \\
&-&\frac{5}{96} i \left(\nabla ^B \overline{R}
\right) \overline{W}^2 \nabla_{\ \ \dot A}^C W_{ABC} 
- \frac{i}{4} \overline{R} \overline{W}^2 \nabla^B 
\nabla_{\ \ \dot A}^C W_{ABC} \nonumber \\
&-& \frac{i}{4} \overline{R} \left( \nabla_{\ \ \dot A}^C \overline{W}^2
\right) \nabla^B W_{ABC} 
- \frac{1}{4} \overline{R} \overline{W}^2 G_{\ \ \dot A}^C \nabla^B W_{ABC} 
\nonumber \\
&-& \frac{5}{12} \overline{R} \overline{W}^2
\left( \nabla^B G_{\ \ \dot A}^C \right) W_{ABC} 
-\frac{1}{2} \left( \nabla^B \nabla_{\underline{\dot A}} G_{\ \
\underline{\dot D}}^C \right) \left( \nabla^{\dot D} \overline{W}^2 \right)
W_{ABC} \nonumber \\
&-& i \left( \nabla _{\ \ \underline{\dot A}}^B G_{\ \ \underline{\dot D}}^C 
\right) \left( \nabla^{\dot D} \overline{W}^2 \right) W_{ABC} 
- \frac{5}{2} \overline{W}^2 G^{B\dot D} \left( \nabla_{\underline{\dot A}} 
G_{\ \ \underline{\dot D}}^C \right) W_{ABC} \nonumber \\
&-& \frac{5}{2} i \left( \nabla^{B\dot D} \overline{W}^2 \right) \left(
\nabla_{\underline{\dot A}} G_{\ \ \underline{\dot D}}^C \right) W_{ABC} 
+ \frac{i}{2} \overline{W}^2 \left( \nabla_{\underline{\dot A}} G_{\ \
\underline{\dot D}}^C \right) \nabla^{B \dot D} W_{ABC} \nonumber \\
&+& \frac{1}{4} \left(\overline{\nabla}^2 \overline{W}^2 \right) \left( 
\nabla^B G_{\ \ \dot A}^C \right) W_{ABC} 
+ \left(\nabla^{\dot D} \nabla_{\ \ \underline{\dot A}}^B \nabla_{\ \
\underline{\dot D}}^C \overline{W}^2 \right) W_{ABC} \nonumber \\
&+& \left(\nabla^{\dot D} \overline{W}^2 \right) \nabla_{\ \ 
\underline{\dot A}}^B \nabla_{\ \ \underline{\dot D}}^C W_{ABC} 
+ \frac{3}{2} \left( \nabla^{\dot D} \nabla_{\ \ \dot D}^C \overline{W}^2 
\right) \nabla_{\ \ \dot A}^B W_{ABC} \nonumber \\
&-& \left( \nabla_{\underline{\dot A}} \nabla_{\ \underline{\dot D}}^C
\overline{W}^2 \right) \nabla^{B\dot D} W_{ABC} 
-2i \overline{W}^2 W_A^{\ \ BD} \nabla_{\ \ \dot A}^C W_{ABC} \nonumber \\
&+& \frac{1}{8} \left( \nabla_{\dot A}
\overline{W}^2 \right) W_A^{\ \ BC} \nabla ^D W_{DBC} 
+\frac{1}{48} \left( \nabla_{\dot A} \overline{W}^2 \right) \left( \nabla^B 
\overline{R} \right) \nabla^C W_{ABC}
\end{eqnarray}

\begin{eqnarray}
&\left. \right.&
\left( \nabla_{\dot A} \nabla^B - \nabla^B \nabla_{\dot A} \right)
\nabla^{\dot D} \nabla_{\ \ \dot D}^C \left( W_{ABC} \overline{W}^2 \right)
\nonumber \\  
&=& \frac{5}{24} \overline{W}^2 \left( \nabla^B \nabla_{\ \ \dot A}^C
\overline{R} \right) W_{ABC} 
+ \frac{1}{12} \overline{W}^2 \left( \nabla_{\
\ \dot A}^C \overline{R} \right) \nabla^B W_{ABC} \nonumber \\
&-& \frac{55}{48} i \overline{W}^2 G_{\ \ \dot A}^C 
\left( \nabla^B \overline{R} \right) W_{ABC} 
+ \frac{3}{8} \left( \nabla^B \overline{R} \right) 
\left( \nabla_{\ \ \dot A}^C \overline{W}^2 \right) W_{ABC} \nonumber \\
&+& \frac{11}{24} \left( \nabla^B
\overline{R} \right) \overline{W}^2 \nabla_{\ \ \dot A}^C W_{ABC} 
-\frac{5}{48} i \left( \nabla_{\dot A} \overline{W}^2 \right) \left(
\nabla^C \overline{R} \right) \nabla^B W_{ABC} \nonumber \\
&+& \frac{1}{2} \overline{R} \left( \nabla_{\ \ \dot A}^C \overline{W}^2 
\right) \nabla^B W_{ABC} 
- \frac{5}{6} i \overline{R} \overline{W}^2 \left(\nabla ^B G_{\ \ \dot A}^C
\right) W_{ABC} \nonumber \\
&-& \frac{5}{6} i \overline{R} \overline{W}^2 G_{\ \ \dot
A}^C \nabla^B W_{ABC} 
+ \frac{1}{3} \overline{R} \overline{W}^2 \nabla^B \nabla_{\ \ \dot A}^C 
W_{ABC} \nonumber \\
&-& 2i \left( \nabla^B \nabla_{\underline{\dot A}} 
G_{\ \ \underline{\dot D}}^C 
\right) \left( \nabla^{\dot D} \overline{W}^2 \right) W_{ABC} 
-6 \left( \nabla_{\ \ \underline{\dot A}}^B
G_{\ \ \underline{\dot D}}^C \right) \left( \nabla^{\dot D} \overline{W}^2
\right) W_{ABC} \nonumber \\
&+& 8 \left( \nabla^{B \dot D} \overline{W}^2 \right) \left(
\nabla_{\underline{\dot A}} G_{\ \ \underline{\dot D}}^C \right) W_{ABC} 
- \frac{3}{2} i \left( \overline{\nabla}^2 \overline{W}^2 \right) \left(
\nabla^B G_{\ \ \dot A}^C \right) W_{ABC} \nonumber \\
&+&5i \left( \nabla^{\dot D}
\overline{W}^2 \right) \left( \nabla_{\underline{\dot A}} G_{\ \
\underline{\dot D}}^C \right) \nabla^B W_{ABC} 
+ 6 G^{B\dot D} \left( \nabla_{\underline{\dot A}} \overline{W}^2 \right) 
\nabla_{\ \ \underline{\dot D}}^C W_{ABC} \nonumber \\
&-& 6 G_{\ \ \dot A}^B \left( \nabla^{\dot D}
\overline{W}^2 \right) \nabla_{\ \ \dot D}^C W_{ABC} 
+10 G^{B\dot D} \left( \nabla_{\underline{\dot A}} 
\nabla_{\ \ \underline{\dot D}}^C \overline{W}^2 \right) W_{ABC} \nonumber \\
&-& 4 G_{\ \ \dot A}^B \left( \nabla^{\dot D}
\nabla_{\ \ \dot D}^C \overline{W}^2 \right) W_{ABC} 
+ \left( \overline{\nabla}^2 \nabla_{\ \ \dot A}^B \overline{W}^2 \right) 
\nabla^C W_{ABC} \nonumber \\
&+& i \left( \nabla^{\dot D} \nabla_{\ \ \dot D}^C \overline{W}^2
\right) \nabla_{\ \ \dot A}^B W_{ABC} 
- \frac{5}{4} i \left( \nabla_{\dot
A} \overline{W}^2 \right) W_A^{\ \ BC} \nabla^D W_{DBC} \nonumber \\
&+& 2i \left( \nabla_{\underline{\dot A}} \nabla_{\ \ \underline{\dot D}}^C
\overline{W}^2 \right) \nabla^{B\dot D} W_{ABC} 
+ \left( \overline{\nabla}^2 \overline{W}^2 \right) \nabla^B 
\nabla_{\ \ \dot A}^C W_{ABC} 
-2 W^2 \nabla_{A\dot A} \overline{W}^2 \nonumber \\
&-& 2i \left(\nabla^{\dot D} \overline{W}^2 \right) \nabla_{\ \
\underline{\dot A}}^B \nabla_{\ \ \underline{\dot D}}^C W_{ABC} 
+ 2i \left( \nabla^{\dot D} \nabla_{\ \ \underline{\dot A}}^B 
\nabla_{\ \ \underline{\dot D}}^C \overline{W}^2 \right) W_{ABC}
\end{eqnarray}


\section{Exact computations to first order in $\alpha$}
\setcounter{equation}{0}
\label{d}
\indent

Here we present the expansion, to first order in $\alpha$, of all the 
terms of (\ref{g2a}):
\begin{equation}
\nabla_{\ \ \dot A}^C \overline{R} = 6 \alpha 
\left(\nabla_{\ \ \dot A}^C W^2 \right) \overline{\nabla}^2 
\overline{W}^2 + 6 \alpha W^2 \overline{\nabla}^2
\nabla_{\ \ \dot A}^C \overline{W}^2 +{\cal O}\left( \alpha ^2\right)
\end{equation}

\begin{eqnarray}
\nabla^{\underline{B}}\nabla_{\ \ \dot A}^{\underline{C}}\overline{R} &=& 
6 \alpha \left( \nabla^{\underline{B}} \nabla_{\ \ \dot A}^{\underline{C}} 
W^2 \right) \overline{\nabla}^2 \overline{W}^2 
+24 i \alpha \left( \nabla_{\ \ \dot A}^{\underline{C}} W^2 \right) 
\nabla^{\dot C} \nabla_{\ \ \dot C}^{\underline{B}} \overline{W}^2
\nonumber \\
&+& 6 \alpha \left( \nabla^{\underline{B}} W^2 \right) \overline{\nabla}^2 
\nabla_{\ \ \dot A}^{\underline{C}} \overline{W}^2 
+ 24 i \alpha W^2 \nabla^{\dot C} \nabla_{\ \ \dot C}^{\underline{B}} 
\nabla_{\ \ \dot A}^{\underline{C}} \overline{W}^2
\nonumber \\
&-& 48 \alpha W^2 W^{BCD} \nabla_{D \dot A} \overline{W}^2 
+{\cal O}\left( \alpha^2\right)
\end{eqnarray}

\begin{eqnarray}
\nabla^{\underline{B}} G_{\ \ \dot A}^{\underline{C}}&=&i \alpha \overline{W}^2
\nabla^{\underline{B}} \nabla_{\ \ \dot A}^{\underline{C}} W^2 
+ i \alpha \left(\nabla^{\underline{B}} \overline{W}^2 \right) 
\nabla_{\ \ \dot A}^{\underline{C}} W^2
\nonumber \\
&+&\frac{3}{2}i \alpha \left( \nabla^{E \dot E} \nabla_{E \dot E} 
\nabla_{D \dot A} \overline{W}^2 \right) W^{BCD} \nonumber \\
&+&3i \alpha W^F_{\ \ DE} \left( \nabla_{F \dot A} \overline{W}^2 \right)
\nabla^{\underline{B}} W^{\underline{C} \underline{D} \underline{E}}
\nonumber \\
&+& 3i \alpha W_{\dot A}^{\ \ \dot E \dot F} \left(\nabla_{\dot E} 
\nabla_{D \dot F} \overline{W}^2 \right) W^{BCD} 
+\frac{3}{4} \alpha W_{\dot A \dot B \dot C} \nabla^2 
\nabla^{\underline{B} \dot B} \nabla^{\underline{C} \dot C} W^2
\nonumber \\
&+& \frac{3}{2} \alpha \left(\nabla^{\dot C} 
\nabla_{D \dot C} \nabla_{E \dot A} \overline{W}^2 \right)  
\nabla^{\underline{B}} W^{\underline{C} \underline{D} \underline{E}}
+{\cal O}\left( \alpha^2\right)
\end{eqnarray}

\begin{eqnarray}
\nabla^{\underline{B}} \nabla_{\underline{\dot A}} 
G_{\ \ \underline{\dot B}}^{\underline{C}}&=&
-i \alpha \left( \nabla^{\underline{B}} W^2 \right)
\nabla_{\underline{\dot A}} \nabla_{\ \ \underline{\dot B}}^{\underline{C}}
\overline{W}^2 -2 \alpha W^2 \nabla_{\ \ \underline{\dot A}}^{\underline{B}}
\nabla_{\ \ \underline{\dot B}}^{\underline{C}} \overline{W}^2
\nonumber \\ 
&-& 2\alpha \left(\nabla_{\ \ \underline{\dot A}}^{\underline{B}} W^2 \right)
\nabla_{\ \ \underline{\dot B}}^{\underline{C}} \overline{W}^2
+ i \alpha \left(\nabla_{\underline{\dot A}} \overline{W}^2 \right)
\nabla^{\underline{B}} \nabla_{\ \ \underline{\dot B}}^{\underline{C}} W^2
\nonumber \\
&+&\frac{3}{2}i \alpha W_{\dot A \dot B \dot C} \nabla^{\underline{B}} 
\nabla^{E \dot E} \nabla_{E \dot E} \nabla^{\underline{C} \dot C} W^2 
\nonumber \\
&-&\frac{3}{4} \alpha \left( \nabla^2 \nabla^{\underline{B} \dot D} 
\nabla^{\underline{C} \dot C} W^2 \right) \nabla_{\underline{\dot A}}
W_{\underline{\dot B} \underline{\dot C} \underline{\dot D}}
\nonumber \\
&+& 3 i \alpha \left( \nabla^D \nabla^{\ \ \dot D}_D
\nabla^{\underline{C} \dot C} W^2 \right)
\nabla_{\ \ \underline{\dot A}}^{\underline{B}}
W_{\underline{\dot B} \underline{\dot C} \underline{\dot D}}
\nonumber \\
&+& \frac{3}{4} \alpha \left(\overline{\nabla}^2 \nabla_{D \underline{\dot B}} 
\nabla_{E \underline{\dot A}} \overline{W}^2 \right)
\nabla^{\underline{B}} W^{\underline{C} \underline{D} \underline{E}}
\nonumber \\
&-& 3 i \alpha W^{\underline{C} DE}
\nabla^{\dot C} \nabla_{\ \ \dot C}^{\underline{B}}
\nabla_{D \underline{\dot B}} \nabla_{E \underline{\dot A}} \overline{W}^2
\nonumber \\
&-&\frac{3}{2} \alpha W^{BCD} W^{\dot D}_{\ \ \dot A \dot B}
\overline{\nabla}^2 \nabla_{D \dot D} \overline{W}^2
\nonumber \\
&-& 12 \alpha W^{\underline{\underline{B}}DE} 
W^{\underline{\underline{C}}}_{\ \ EF} \nabla_{\underline{D} 
\underline{\dot A}} \nabla_{\ \ \underline{\dot B}}^{\underline{F}}
\overline{W}^2 
\nonumber \\
&+& 3 i \alpha W^{BCD} \left(\nabla^{\dot C} 
\nabla_D^{\ \ \dot D} \overline{W}^2 \right) \nabla_{\underline{\dot A}} 
W_{\underline{\dot B} \underline{\dot C} \underline{\dot D}}
\nonumber \\
&-& 3 i \alpha 
W_{\dot A \dot B \dot C} \left( \nabla_D \nabla_E^{\ \ \dot C} W^2 \right)
\nabla^{\underline{B}} W^{\underline{C} \underline{D} \underline{E}}
\nonumber \\
&+&\frac{3}{2}i \alpha W_{\dot A \dot B \dot C} W^{BCD}
\nabla^2 \nabla_D^{\ \ \dot C} W^2 
\nonumber \\
&-& 6 \alpha \left(\nabla^{\underline{B}}_{\ \ \underline{\dot A}} 
W_{\underline{\dot B} \underline{\dot C} \underline{\dot D}} \right) 
W^{\dot C \dot D \dot E} \nabla^{\underline{C}}_{\ \ \dot E} W^2
\nonumber \\
&+& 3 i \alpha \left( \nabla_{\underline{\dot A}} 
W_{\underline{\dot B} \underline{\dot C} \underline{\dot D}} \right)
W^{\dot C \dot D \dot E}
\nabla^{\underline{B}} \nabla^{\underline{C}}_{\ \ \dot E} W^2
+ {\cal O}\left( \alpha^2\right)
\end{eqnarray}

\begin{eqnarray}
\nabla^{\underline{B}}_{\ \ \underline{\dot A}} 
G_{\ \ \underline{\dot B}}^{\underline{C}}&=& -\alpha
\left(\nabla_{\underline{\dot A}} \overline{W}^2 \right)
\nabla^{\underline{B}} \nabla_{\ \ \underline{\dot B}}^{\underline{C}} W^2
-i \alpha W^2 \nabla_{\ \ \underline{\dot A}}^{\underline{B}}
\nabla_{\ \ \underline{\dot B}}^ {\underline{C}} \overline{W}^2  
\nonumber \\  
&-&\frac{3}{2} \alpha \left(\nabla^{\dot C} \nabla_{D \dot C}
\nabla_{E \underline{\dot B}} \overline{W}^2 \right) 
\nabla^{\underline{B}}_{\ \ \underline{\dot A}} 
W^{\underline{C} \underline{D} \underline{E}}
\nonumber \\
&+&\frac{3}{2} \alpha W^{\underline{C} EF}
\nabla^{\dot C} \nabla_{\ \ \dot C}^{\underline{B}}
\nabla_{E \underline{\dot A}} \nabla_{F \underline{\dot B}} \overline{W}^2 
\nonumber \\
&+&\frac{3}{4} \alpha W^{BCD} \nabla_{\underline{\dot A}} \nabla^{E \dot E} 
\nabla_{E \dot E} \nabla_{D \underline{\dot B}} \overline{W}^2 
\nonumber \\
&+&\frac{3}{2} \alpha \left(\nabla_{\underline{\dot A}} 
\nabla_{\ \ \underline{\dot B}}^F \overline{W}^2 \right) W_{DEF}
\nabla^{\underline{B}} W^{\underline{C} \underline{D} \underline{E}}
\nonumber \\
&+& 3i \alpha \left(\nabla_{\ \ \underline{\dot B}}^F \overline{W}^2 \right) 
W_{DEF} \nabla^{\underline{B}}_{\ \ \underline{\dot A}}
W^{\underline{C} \underline{D} \underline{E}}
\nonumber \\
&-& 6i \alpha W^{\underline{\underline{B}}DE} 
W^{\underline{\underline{C}}}_{\ \ EF} \nabla_{\underline{D} 
\underline{\dot A}} \nabla_{\ \ \underline{\dot B}}^{\underline{F}}
\overline{W}^2 
+\left. \mathrm{h.c.}\right. +{\cal O}\left( \alpha^2\right)
\end{eqnarray}


\end{document}